\documentclass[12pt,a4paper]{article}
\usepackage[utf8]{inputenc}
\usepackage{epsfig}
\usepackage{mathtools}
\usepackage{amsfonts}
\usepackage{amssymb}
\usepackage{amsmath}
\usepackage{cancel}
\usepackage{color}
\usepackage{slashed}
\usepackage{cite}
\usepackage{caption}
\usepackage{subcaption}
\usepackage{comment}
\usepackage{marginnote}

\usepackage{tikz}
\usetikzlibrary{shapes}
\usetikzlibrary{trees}
\usetikzlibrary{calc}
\usetikzlibrary{decorations.pathmorphing}
\usetikzlibrary{decorations.markings}

\allowdisplaybreaks

\def\bP{{\bf P}}

\def\OO{\mathcal{O}}
\def\Z{\mathcal{Z}}
\def\P {{\bf P}}
\def\Pt{\tilde{\bf P}}
\def\p {{\bf p}}
\def\pt{\tilde{\bf p}}
\def\Q {{\bf Q}}
\def\Qt{\tilde{\bf Q}}

\newcommand{\sign}{\text{sign}}

\def\N{\mathcal{N}}

\def\PP{\mathcal{P}}

\def\j{\bullet}
\def\wt0{\tilde{w}_0}

\def\Ft{\widetilde{F}}
\def\H{\mathbb{H}}
\def\PP{\mathbb{P}}

\begin{document}

\begin{center}

\vspace{1cm}

{ \Large\bf Anomalous dimensions of twist 2 operators \\
	and $\N=4$ SYM quantum spectral curve} \vspace{1cm}

{\large Andrei I. Onishchenko$^{1,2,3}$}\vspace{0.5cm}

{\it 
$^1$Bogoliubov Laboratory of Theoretical Physics, Joint
Institute for Nuclear Research, Dubna, Russia, \\
$^2$Skobeltsyn Institute of Nuclear Physics, Moscow State University, Moscow, Russia \\
$^3$ Budker Institute of Nuclear Physics, Novosibirsk, Russia
}\vspace{1cm}

\abstract{
We present algorithmic perturbative solution of $\N=4$ SYM quantum spectral curve in the case of twist 2 operators, valid to in principle arbitrary order in coupling constant. The latter treats operator spins as arbitrary integer values and is written in terms of special class of functions -- products of rational functions in spectral parameter with sums of Baxter polynomials and Hurwitz functions. It is shown that this class of functions is closed under elementary operations, such as shifts, partial fractions, multiplication by spectral parameter and differentiation. Also, it is fully sufficient to solve arising non-homogeneous  multiloop Baxter and first order difference equations. As an application of the proposed method we present the computation of anomalous dimensions of twist 2 operators up to four loop order.}

\end{center}

\begin{center}
	Keywords: quantum spectral curve, spin chains, anomalous dimensions, \\ $\N=4$ SYM, Baxter equations
\end{center}

\newpage

\tableofcontents{}\vspace{0.5cm}

\renewcommand{\theequation}{\thesection.\arabic{equation}}

\section{Introduction}

Recently, we have seen a lot of progress in the study of integrability in quantum field theories with extended supersymmetry, for a review and introduction see \cite{IntegrabilityReview,IntegrabilityPrimer,IntegrabilityDeformations,IntegrabilityDefects,IntegrabilityStructureConstants,FishnetCFTreview}. In particular a lot of attention was payed to the study of anomalous dimensions of twist 2 operators in $\N=4$ super Yang-Mills (SYM) theory together with related DGLAP and BFKL evolution dynamics, see \cite{DGLAP-BFKL-review,N4SYM-BFKL-QSC-review} for a review. The latter could be conveniently described in a language of integrable long-range super spin chains,  where calculation of anomalous dimensions reduces to calculation of spin chains spectra. The most impressive results in such type of calculations  were achieved within quantum spectral curve (QSC) approach, see \cite{IntroductionQSC,LevkovichReview} for an introduction.  QSC is basically a set of algebraic relations for Baxter type $Q$-functions together with their analyticity and Riemann-Hilbert monodromy conditions \cite{N4SYMQSC1,N4SYMQSC2,twistedN4SYMQSC,N4SYMQSC3,N4SYMQSC4,ABJMQSC,ABJMQSCdetailed,QSCetadeformed}. Within  quantum spectral curve approach one can relatively easy obtain numerical solution for any coupling and spin chain state \cite{QSCnumericsN4SYM1,QSCnumericsN4SYM2,QSCnumericsABJM}. Also, QSC formulation allows to construct analytic perturbative solutions at weak coupling up to, in principle, arbitrary loop order \cite{VolinPerturbativeSolution,ABJMQSC12loops,N4SYMQSC4}. These analytical solutions, in contrast to numerical ones \cite{QSCnumericsN4SYM1,QSCnumericsN4SYM2,QSCnumericsABJM},  are however limited to the situation when the states quantum numbers are given explicitly by some fixed integers.

The aim of the present article is to provide analytical perturbative  solution of $\N=4$ SYM quantum spectral curve equations in the case of twist 2 operators treating operator spins as arbitrary integer values. The obtained solution is valid to in principle arbitrary order in coupling constant. Before such a solution for arbitrary spins was considered  only for asymptotic multiloop Baxter equation \cite{Kotikov1,Kotikov2} and as such do not account for finite size wrapping corrections.  The authors of \cite{Kotikov1,Kotikov2} used Mellin space technique and were able to obtain at most four-loop asymptotic contribution to anomalous dimensions of twist 2 operators. Our own experience with Mellin space technique in the context of quantum spectral curve \cite{ABJM_QSC_Mellin} has also showed that the use of this technique becomes very complex at higher loop orders. The approach we are going to present here on the other hand does all calculations directly in spectral parameter space and gives algorithmic perturbative solution at any loop order. It is further development of the technique originally developed for twist 1 operators in ABJM model in \cite{uspaceTMF,ABJMtwist1}. Similar to \cite{uspaceTMF,ABJMtwist1}, the presented solution is based on the existence of a class of functions - products of rational functions in spectral parameter with sums of Baxter polynomials and Hurwitz functions, which is closed under elementary operations, such as shifts and partial fractions,  as well as differentiation. The introduced class of function is sufficient for finding solutions of the involved inhomogeneous Baxter and first order difference equations using recursive construction of the dictionary of particular solutions for canonical types of inhomogeneities. 
The presented approach  has the potential for generalizations to higher twists of operators, as well as to other theories such as ABJM model and twisted $\mathcal{N}=4$ SYM and ABJM models.  Moreover, similar ideas should be also applicable to the study of BFKL regime within QSC approach \cite{GromovBFKL1,GromovBFKL2,BFKLnonzeroConformalSpin}.

This paper is organized as follows.  In the next section we
give necessary details on $\N=4$ SYM quantum spectral curve equations with the emphasis on $\P\mu$-system. Section \ref{Riemann-Hilbert-solution} contains the details of the solution of Riemann-Hilbert problem for $\P\mu$ - system relevant to the calculation of anomalous dimensions of twist 2 operators. Next, section \ref{solution-difference-eqns} discusses the  solution of the involved inhomogeneous difference equations. Section \ref{constraintsSolutionAnomDimensions} gives details on the solution of the constraint equations. Finally, in section \ref{Conclusion} we come with our conclusion.  Appendices  contain details of our calculation dropped in the main body of the paper together with notation used.

\section{$\N=4$ SYM quantum spectral curve}\label{N4SYM-QSC}

At present the most advanced framework for the calculation of anomalous dimensions of different operators in $\N=4$ SYM is provided by quantum spectral curve (QSC) approach. The latter is the reformulation of Thermodynamic Bethe Ansatz (TBA) equations for long-range $\mathfrak{psu} (2,2|4)$ spin chain \cite{TBAN4,TBAN4proposal,TBAexcitedstates,TBAMirrorModel,YsystemAdS5,TBAfromYsystem,WronskianSolution,SolvingYsystem} in terms of a finite set of functional equations ($Q$-system) together with their analyticity and Riemann-Hilbert monodromy conditions \cite{N4SYMQSC1,N4SYMQSC2,twistedN4SYMQSC,N4SYMQSC3,N4SYMQSC4}. In the $\mathfrak{sl}(2)$ sector we are interested in the QSC equations (so called $\P\mu$-system) take the form of Riemann-Hilbert problem \cite{N4SYMQSC1,N4SYMQSC2}:
\begin{align}
\mu_{a b} -\tilde{\mu}_{a b} &= \tilde{\P}_a\P_b - \tilde{\P}_b\P_a\, ,\label{eq:Pmu1} \\
\tilde{\P}_a &= (\mu\chi)_a^{~b} \P_b\, , \label{eq:Pmu2} \\
\tilde{\mu}_{a b} &= \mu_{a b}^{[2]}\, , \label{eq:Pmu3}
\end{align}
where $(\mu\chi)_a^{~b}\equiv \mu_{a c}\chi^{c b}$ and
\begin{equation}
\chi^{a b} = \begin{pmatrix}
0 & 0 & 0 & -1 \\ 0 & 0 & +1 & 0 \\ 0 & -1 & 0 & 0 \\
+1 & 0 & 0 & 0
\end{pmatrix}\, 
\end{equation}
Here, indexes $a, b$ take values $1, 2, 3, 4$ and $\mu_{a b}$ is antisymmetric matrix, which entries
\begin{equation}
\mu_{12}=\mu_{1},\quad\mu_{13}=\mu_{2},\quad\mu_{14}=\mu_{23}=\mu_{3},\quad\mu_{24}=\mu_{4},\quad\mu_{34}=\mu_{5}.
\end{equation}
are subject to constraint
\begin{equation}
\mu_{1}\mu_{5} - \mu_{2}\mu_{4} + \mu_{3}^2 = 1.
\end{equation}
The functions $\P_a$ and $\mu_{a b}$ are functions of spectral parameter $u$.
The $\P_a$ - functions have only one Zhukovsky cut\footnote{Here and below $g^2 =\frac{N_c g_{YM}^2}{16\pi^2}$, where $g_{YM}$ is $\N=4$ SYM coupling constant and $N_c$ is the number of colors.} ($u\in [-2g, 2g]$) on the defining (physical) Riemann sheet, while functions $\mu_{a b}$ have infinitely many branch points at positions $u = \pm 2 g + i\mathbb{Z}$. Next, $f^{[n]}(u) = f(u+\frac{i n}{2})$ and $\tilde{f}(u)$ denotes analytic continuation of function $f (u)$ around one of its branch points on the real axis located at $-2g$ and $2g$. The boundary conditions for the above Riemann-Hilbert problem are specified by the large $u$ asymptotics of $\P_a$ and $\mu_{i}$ functions:
\begin{align}
& \P_1 \simeq A_1 u^{-\frac{L+2}{2}}\, , \P_2 \simeq A_2 u^{-\frac{L}{2}}\, , \P_3\simeq A_3 u^{\frac{L-2}{2}}\, , \P_4\simeq A_4 u^{\frac{L}{2}}\, , \nonumber \\
& \mu_{1}\sim u^{\Delta - L}, \,  \mu_{2}\sim u^{\Delta - 1}, \, \mu_{3}\sim u^{\Delta}, \,  \mu_{4}\sim u^{\Delta + L} , \,  \mu_{5}\sim u^{\Delta + L}
\end{align}
with
\begin{align}
A_1 A_4 &= \frac{[(L-S+2)^2 - \Delta^2][(L+S)^2 - \Delta^2]}{16 i L (L+1)}\, , \nonumber \\
A_2 A_3 &= \frac{[(L+S-2)^2-\Delta^2][(L-S)^2 - \Delta^2]}{16 i L (L-1)}\, . \label{eq:A3A4}
\end{align}
Here $L$ is the twist, $S$ - spin and $\Delta$ is conformal dimension of the $\mathfrak{sl}(2)$ operator under consideration. Finally, the operator anomalous dimension $\gamma$ is related to its conformal dimension as $\gamma = \Delta - S - L$.

\section{Riemann-Hilbert problem for $\bP\mu$-system}\label{Riemann-Hilbert-solution}

It is convenient to start the solution of the original Riemann-Hilbert problem \eqref{eq:Pmu1}-\eqref{eq:Pmu3} by writing  down its defining equations for all values of indexes $a$ and $b$. Then, their straightforward algebraic solution shows that all functions $\mu_i$ and $\Pt_a$ can be expressed in terms of $\mu_1$, $\mu_4$, $\Pt_2$ and $\P_a$ - functions. Introducing notation 
\begin{equation}
q_{1}=g^{L}\mu_{1}^{[1]},\quad q_{4}=g^{L}\mu_{4}^{[1]},\quad r_{2}=g^{L}\frac{\Pt_{2}^{[-1]}}{\P_{2}^{[-1]}} \label{eq:q1q4r2}
\end{equation}
we have 
\begin{align}
g^{L}\mu_{2} & =\frac{1}{\left(\P_{2}\right)^{2}}q_{1}^{[1]}+\frac{\left(\P_{1}\right)^{2}}{\left(\P_{2}\right)^{2}}q_{4}^{[-1]}-\frac{2\P_{1}}{\P_{2}}r_{2}^{[1]}+\left(\frac{\P_{3}}{\P_{2}}-\frac{1}{\left(\P_{2}\right)^{2}}+\frac{\P_{1}\P_{4}}{\left(\P_{2}\right)^{2}}\right)q_{1}^{[-1]}\,,\\
g^{L}\mu_{2}^{[2]} & =-\frac{1}{\left(\P_{2}\right)^{2}}q_{1}^{[-1]}+\frac{\left(\P_{1}\right)^{2}}{\left(\P_{2}\right)^{2}}q_{4}^{[1]}-\frac{2\P_{1}}{\P_{2}}r_{2}^{[1]}+\left(\frac{\P_{3}}{\P_{2}}+\frac{1}{\left(\P_{2}\right)^{2}}+\frac{\P_{1}\P_{4}}{\left(\P_{2}\right)^{2}}\right)q_{1}^{[1]}\,,\\
g^{L}\mu_{3} & =-r_{2}^{[1]}+\frac{\P_{1}}{\P_{2}}q_{4}^{[-1]}+\frac{\P_{4}}{\P_{2}}q_{1}^{[-1]}\,,\\
g^{L}\mu_{3}^{[2]} & =-r_{2}^{[1]}+\frac{\P_{1}}{\P_{2}}q_{4}^{[1]}+\frac{\P_{4}}{\P_{2}}q_{1}^{[1]}\,,\\
g^{L}\mu_{5} & =\frac{1}{\left(\P_{2}\right)^{2}}q_{4}^{[1]}+\frac{2\P_{4}}{\P_{2}}r_{2}^{[1]}-\frac{\left(\P_{4}\right)^{2}}{(\P_{2})^{2}}q_{1}^{[-1]}+\left(\frac{\P_{3}}{\P_{2}}-\frac{1}{\left(\P_{2}\right)^{2}}-\frac{\P_{1}\P_{4}}{\left(\P_{2}\right)^{2}}\right)q_{4}^{[-1]}\:,\\
g^{L}\mu_{5}^{[2]} & =-\frac{1}{\left(\P_{2}\right)^{2}}q_{4}^{[-1]}+\frac{2\P_{4}}{\P_{2}}r_{2}^{[1]}-\frac{\left(\P_{4}\right)^{2}}{(\P_{2})^{2}}q_{1}^{[1]}+\left(\frac{\P_{3}}{\P_{2}}+\frac{1}{\left(\P_{2}\right)^{2}}-\frac{\P_{1}\P_{4}}{\left(\P_{2}\right)^{2}}\right)q_{4}^{[1]}
\end{align}
and 
\begin{align}
g^{L}\Pt_{1} & =\P_{1}r_{2}^{[1]}+\frac{1}{\P_{2}}q_{1}^{[-1]}-\frac{1}{\P_{2}}q_{1}^{[1]}\:,\label{eq:Pt1}\\
g^{L}\Pt_{3} & =-\frac{\P_{1}}{\left(\P_{2}\right)^{2}}q_{4}^{[-1]}+\frac{\P_{1}}{\left(\P_{2}\right)^{2}}q_{4}^{[1]}+\P_{3}r_{2}^{[1]}-\frac{\P_{4}}{\left(\P_{2}\right)^{2}}q_{1}^{[-1]}+\frac{\P_{4}}{\left(\P_{2}\right)^{2}}q_{1}^{[1]}\:,\label{eq:Pt3}\\
g^{L}\Pt_{4} & =-\frac{1}{\P_{2}}q_{4}^{[-1]}+\frac{1}{\P_{2}}q_{4}^{[1]}+\P_{4}r_{2}^{[1]} \label{eq:Pt4}
\end{align}  
The same solution for $\mu_1$, $\mu_4$, $\Pt_2$ - functions gives us the following difference equations:
\begin{multline}
\frac{1}{\left(\P_{2}^{[1]}\right)^{2}}q_{1}^{[2]}+\frac{1}{\left(\P_{2}^{[-1]}\right)^{2}}q_{1}^{[-2]}\\
+\left[\frac{1}{\P_{2}^{[1]}\P_{2}^{[-1]}}\left(\P_{1}^{[1]}\P_{4}^{[-1]}-\P_{1}^{[-1]}\P_{4}^{[1]}\right)+\frac{\P_{3}^{[1]}}{\P_{2}^{[1]}}-\frac{\P_{3}^{[-1]}}{\P_{2}^{[-1]}}-\frac{1}{\left(\P_{2}^{[1]}\right)^{2}}-\frac{1}{\left(\P_{2}^{[-1]}\right)^{2}}\right]q_{1}\\
+\left(\frac{\P_{1}^{[-1]}}{\P_{2}^{[-1]}}-\frac{\P_{1}^{[1]}}{\P_{2}^{[1]}}\right)\left(r_{2}+r_{2}^{[2]}\right)=0\:,\label{eq:Baxter1}
\end{multline}

\begin{equation}
r_{2}-r_{2}^{[2]}+\left(\frac{\P_{4}^{[1]}}{\P_{2}^{[1]}}-\frac{\P_{4}^{[-1]}}{\P_{2}^{[-1]}}\right)q_{1}+\left(\frac{\P_{1}^{[1]}}{\P_{2}^{[1]}}-\frac{\P_{1}^{[-1]}}{\P_{2}^{[-1]}}\right)q_{4}=0\, ,\label{eq:firstorder}
\end{equation}

\begin{multline}
\frac{1}{\left(\P_{2}^{[1]}\right)^{2}}q_{4}^{[2]}+\frac{1}{\left(\P_{2}^{[-1]}\right)^{2}}q_{4}^{[-2]}\\
+\left[\frac{1}{\P_{2}^{[1]}\P_{2}^{[-1]}}\left(\P_{1}^{[1]}\P_{4}^{[-1]}-\P_{1}^{[-1]}\P_{4}^{[1]}\right)+\frac{\P_{3}^{[1]}}{\P_{2}^{[1]}}-\frac{\P_{3}^{[-1]}}{\P_{2}^{[-1]}}-\frac{1}{\left(\P_{2}^{[1]}\right)^{2}}-\frac{1}{\left(\P_{2}^{[-1]}\right)^{2}}\right]q_{4}\\
+\left(\frac{\P_{4}^{[1]}}{\P_{2}^{[1]}}-\frac{\P_{4}^{[-1]}}{\P_{2}^{[-1]}}\right)\left(r_{2}+r_{2}^{[2]}\right)=0\:.\label{eq:Baxter2}
\end{multline}
Note, that two Baxter equations above are coupled only though $r_2$ - function. Now, to find a perturbative solution of the obtained equation we will follow the work of \cite{VolinPerturbativeSolution} and employ for $\P_a$ - functions the  weak coupling ansatz\footnote{Here, we fixed residual gauge symmetry of QSC equations by setting $A_1=g^2$ and $A_2=1$ \cite{VolinPerturbativeSolution}} as $\P_{a}\equiv(gx)^{-L/2}\p_{a}$, where
\begin{align}
\p_{1} & =\frac{g}{x}+\sum_{k=2}^{\infty}\sum_{l=0}^{\infty}c_{1,k}^{(l)}\frac{g^{k+2l}}{x^{k}}\,,\label{eq:Pansatz1}\\
\p_{2} & =1+\sum_{k=2}^{\infty}\sum_{l=0}^{\infty}c_{2,k}^{(l)}\frac{g^{k+2l}}{x^{k}}\,,\label{eq:Pansatz2}\\
\p_{3} & =\sum_{l=0}^{\infty}A_{3}^{(l)}g^{2l}u^{L-1}+\sum_{j=1}^{L-2}\sum_{l=0}^{\infty}m_{j}^{(l)}g^{2l}u^{j}+\sum_{k=1}^{\infty}\sum_{l=0}^{\infty}\frac{c_{3,k}^{(l)}g^{k+2l}}{x^{k}}\,,\label{eq:Pansatz3}\\
\p_{4} & =\sum_{l=0}^{\infty}A_{4}^{(l)}g^{2l}u^{L}+\sum_{j=1}^{L-1}\sum_{l=0}^{\infty}n_{j}^{(l)}g^{2l}u^{j}+\sum_{k=2}^{\infty}\sum_{l=0}^{\infty}\frac{c_{4,k}^{(l)}g^{k+2l}}{x^{k}} \label{eq:Pansatz4}
\end{align}
and $x$ stands for Zhukovsky variable $x=\frac{u}{2g}\left(1+\sqrt{1-\frac{4g^{2}}{u^{2}}}\right)$.  The analytical continuation of the ansatz for $\P$-functions  through the cut on real axis is simple and is given by \cite{VolinPerturbativeSolution}:
\begin{equation}
\Pt_a = \left(\frac{x}{g}\right)^{L/2} \pt_a\, , \quad \pt_a = \pt_a\Big|_{x\to 1/x}\, .
\end{equation}
On the other hand the same $\Pt_a$ - functions ($a=1,3,4$) can be obtained through the equations \eqref{eq:Pt1}- \eqref{eq:Pt4}, while the $\Pt_2$ - function can be expressed though the solution for $r_2$ function from Eq. \eqref{eq:firstorder}. These two possible ways of obtaining the same quantities give us means to determine unknown coefficients $c_{a,k}^{(l)}$, $A_3^{(l)}$, $A_4^{(l)}$, $m_j^{(l)}$ and $n_j^{(l)}$. The latter for twist 2 operators (spin chain states) are functions of spin $S$ only. Note, that to find perturbative solution of $\P\mu$ - system we also use perturbative expansion of $q_1$, $q_2$ and $r_2$ functions as ($k=1,4$):
\begin{align}
q_k (u) &= \sum_{l=0}^{\infty} q_k^{(l)}(u) g^{2l}, \\r_2 (u) &= \sum_{l=0}^{\infty} r_2^{(l)}(u) g^{2l}.
\end{align} 
In addition, when solving nonhomogeneous Baxter equations \eqref{eq:Baxter1} and \eqref{eq:Baxter2} we need to specify boundary condition to fix uncertainty in the homogeneous pieces of solutions. This can be conveniently done by requiring that the following combinations of functions\footnote{This follows from the fact, that $\mu_i$ - functions have square root cuts on the real axis and collision of branch points in weak coupling expansion at $u=0$ may produce poles.}
\begin{align}
\mu_{a}+\tilde{\mu}_{a} & =\mu_{a}+\mu_{a}^{[2]}\:, \label{eq:mupoles1}\\
\frac{\mu_{a}-\tilde{\mu}_{a}}{\sqrt{u^{2}-4g^{2}}} & =\frac{\mu_{a}-\mu_{a}^{[2]}}{\sqrt{u^{2}-4g^{2}}} \label{eq:mupoles2}
\end{align}
are free of cuts on the whole real axis. In particular in weak coupling expansions these combinations should be free of poles at $u=0$.

In what follows we will consider perturbative solution of the described Riemann-Hilbert problem in a special case of twist 2 operators, so from now on we set $L=2$.   

\section{Solution of difference equations} \label{solution-difference-eqns}

The most complicated step in the solution of Riemann-Hilbert problem for $\P\mu$ - system described in previous section is the solution of non-homogeneous difference equations \eqref{eq:Baxter1}-\eqref{eq:Baxter2}. The solution start with leading order of Eq. \eqref{eq:Baxter1} and proceeds recursively. At leading order Eq. \eqref{eq:Baxter1} is given by
\begin{equation}
(u+\frac{i}{2})^{2}q(u+i)+(u-\frac{i}{2})^{2}q(u-i)-(2u^{2}-\frac{1}{2}-S(S+1))q(u)=0\label{eq:q1-LO}
\end{equation}
Here, we have accounted that $A_3^{(0)}=-i S (S+1)$ as follows from Eq. \eqref{eq:A3A4} and previously set value of $A_2=1$. The solution of this equation is easy to find\footnote{One can use Mellin transform for example} and is given by 
\begin{equation}
Q_{S}(u) = ~_{3}F_{2}\left(-S,S+1,\frac{1}{2}-i\,u;1,1;1\right)\,\label{eq:Baxter-polynomial-def}
\end{equation}
times some spin dependent constant $\alpha$. Note, that the normalization of our Baxter polynomial is different from \cite{VolinPerturbativeSolution} and is chosen in such a way as to have simple elementary operations for the latter. Now, we claim that the solutions of all other difference equations can be done using techniques similar to \cite{uspaceTMF,ABJMtwist1}. In particular all calculations can be done algebraically within the special class of functions: products of rational functions in spectral parameter with sums of above Baxter polynomials and Hurwitz functions\footnote{See Appendix \ref{Hurwitz-functions} for their definitions.}. Lets see how it is done.

\subsection{Elementary operations}\label{elementaryopSection}

The solution of difference equations presented below relies heavily on our ability to write down their inhomogeneous parts in a prescribed canonical form, such that their particular solutions should be considered only for inhomogeneities of these special types. For that reason we need to know expressions for elementary operations, such as shifts, partial fractions and multiplication by spectral parameter of Baxter polynomial. All of these elementary operations were successfully found empirically. For shifts and multiplication by simple fractions we obtained the following rules\footnote{Spin $S$ is assumed to be even.}:
\begin{align}
Q^{[2]}_{S} & =Q_{S}+2\sum_{k=1}^{S}(-1)^{k}(1+2S-2k)Q_{S-k}\sum_{n=0}^{k-1}\frac{1}{S-n}\,,\\
Q^{[-2]}_{S} & =Q_{S}+2\sum_{k=1}^{S}(1+2S-2k)Q_{S-k}\sum_{n=0}^{k-1}\frac{1}{S-n}
\end{align}     
and 
\begin{align}
\frac{Q_{s}}{u+i/2} & =\frac{1}{u+i/2}-2i\sum_{k=1}^{S}(-1)^{k}(1+2S-2k)Q_{S-k}(u)\sum_{n=0}^{k-1}\frac{(-1)^{n}}{(S-n)^{2}}\,,\\
\frac{Q_{s}}{u-i/2} & =\frac{(-1)^{S}}{u-i/2}+2i\sum_{k=1}^{S}(1+2S-2k)Q_{S-k}(u)\sum_{n=0}^{k-1}\frac{(-1)^{n}}{(S-n)^{2}}
\end{align}
Then for $Q$-sums\footnote{See Appendix \ref{VQandVwQsums} for their definition and notation.} we have
\begin{align}
Q_{S}^{[2\sigma]} =& Q_{S}+2\langle Q|(-\sigma)^{\j}(1+2S-2\j),\frac{1}{S-\j+1}\rangle\nonumber \\ &+2\langle Q|(-\sigma)^{\j}\frac{(1+2S-2\j)}{S-\j+1}\rangle\,,\\
\langle Q|w,W\rangle^{[2\sigma]}  =& \langle Q|w,W\rangle+2\langle Q|(-\sigma)^{\j}(1+2S-2\j),\frac{1}{S-\j+1},(-\sigma)^{\j}w(\j),W\rangle\nonumber \\
& +2\langle Q|(-\sigma)^{\j}\frac{(1+2S-2\j)}{S-\j+1},(-\sigma)^{\j}w(\j),W\rangle\:,
\end{align}
and 
\begin{align}
\frac{Q_{S}}{u+i/2} & =\frac{1}{u+i/2}+2i\langle Q|(-1)^{\j}(1+2S-2\j),\frac{(-1)^{\j}}{(S-\j+1)^{2}}\rangle+2i\langle Q|\frac{(1+2S-2\j)}{(S-\j+1)^{2}}\rangle\,,\\
\frac{Q_{S}}{u-i/2} & =\frac{1}{u-i/2}-2i\langle Q|(1+2S-2\j),\frac{(-1)^{\j}}{(S-\j+1)^{2}}\rangle-2i\langle Q|\frac{(-1)^{\j}(1+2S-2\j)}{(S-\j+1)^{2}}\rangle\,,
\end{align}

\begin{align}
\frac{1}{u+i/2}\langle Q|w,W\rangle & =\frac{1}{u+i/2}\langle w,W\rangle+2i\langle Q|\frac{(1+2S-2\j)}{(S-\j+1)^{2}},w,W\rangle\nonumber \\
& +2i\langle Q|(-1)^{\j}(1+2S-2\j),\frac{(-1)^{\j}}{(S-\j+1)^{2}},w,W\rangle\:,\\
\frac{1}{u-i/2}\langle Q|w,W\rangle & =\frac{1}{u-i/2}\langle(-1)^{\j}w(\j),W\rangle-2i\langle Q|\frac{(-1)^{\j}(1+2S-2\j)}{(S-\j+1)^{2}},(-1)^{\j}w(\j),W\rangle\nonumber \\
& -2i\langle Q|(1+2S-2\j),\frac{(-1)^{\j}}{(S-\j+1)^{2}},(-1)^{\j}w(\j),W\rangle\:.
\end{align}
Here $w$ is some single weight and $W$ denotes any (including null) sequence of other weights in a sum. As we already noted, the above relations are valid only for even spin values. It turns out however that the inhomogeneities of our difference equations will also include $Q_{S-1}(u)$ polynomials. So we are also require similar relations for $Q_{S-1}(u)$ Baxter polynomials. The latter are given by 
\begin{align}
Q_{S-1}^{[2\sigma]} =& Q_{S-1} -2\sigma\langle Q|(-\sigma)^{\j}(1+2S-2\j),\frac{1}{S-\j+1}\rangle \nonumber \\
& +2\sigma\langle Q|\frac{(-\sigma)^{\j}}{S-\j+1}\rangle + \frac{2\sigma}{S}\langle Q|(-\sigma)^{\j}(1-2\j)\rangle\, ,
\end{align}
and 
\begin{align}
\frac{1}{u+i/2}Q_{S-1} =& \frac{1}{u+i/2} + 2i\langle Q|((-1)^{\j}1+2S-2\j), \frac{(-1)^{\j}}{(S-\j+1)^2}\rangle \nonumber\\ 
& +4i\langle Q|\frac{1}{S-\j+1}\rangle - 2i\langle Q|\frac{1}{(S-\j+1)^2}\rangle + \frac{2i}{S^2}\langle Q|(-1)^{\j}(1+2S-2\j)\rangle\, , \\
\frac{1}{u-i/2}Q_{S-1} =& -\frac{1}{u-i/2} + 2i\langle Q|(1+2S-2\j), \frac{(-1)^{\j}}{(S-\j+1)^2}\rangle \nonumber\\ 
& +4i\langle Q|\frac{(-1)^{\j}}{S-\j+1}\rangle - 2i\langle Q|\frac{(-1)^{\j}}{(S-\j+1)^2}\rangle + \frac{2i}{S^2}\langle Q|(1+2S-2\j)\rangle\, .
\end{align}
The next elementary operation which is often used both for the reduction of inhomogeneities to canonical form and finding particular solutions for each type of canonical inhomogeneities is the multiplication of Baxter polynomials by spectral parameter $u$. The latter was found to have the form:
\begin{equation}
u\,Q_{S}(u)=\frac{i}{2}\frac{1}{(2S+1)}(S^{2}Q_{S-1}(u)-(S+1)^{2}Q_{S+1}(u)) \label{eq:umult}
\end{equation}
For $Q$-sums it is then given by\footnote{See Appendix \ref{VQandVwQsums} for definition and notation for $Q^{\{n\}}$ - sums}
\begin{align}
u\langle Q|w,W\rangle =& u\, w(S+1/2)\langle Q|1,W\rangle 
+\frac{i}{2}\langle Q^{\{-1\}}|\frac{w(\j)-w(S+1/2)}{2S-2j+1},W\rangle \nonumber \\
& -\frac{i}{2}\langle Q^{\{1\}}|\frac{w(\j)-w(S+1/2)}{2S-2j+1},W\rangle\, .
\label{umultVQ}
\end{align}
Here we assumed that the weight $w(\j)$ does not contain factors $(-1)^{\j}$. If it does, then the required rule is obtained with the substitution of $\langle Q|1,W\rangle$ with  $\langle Q|(-1)^{\j},W\rangle$ in the rhs of Eq. \eqref{umultVQ}.  The $Q^{\{n\}}$ - sums can be further reduced to $Q$ - sums, see Appendix \ref{VQandVwQsums} for details. This reduction is in particular based on the rules for the reduction of spin index of Baxter polynomials. The latter were found to be given by
\begin{align}
(S+1)^{2}\left(\begin{array}{c}
Q_{S+1}^{[1]}\\
Q_{S+1}^{[-1]}
\end{array}\right) =& \left(\begin{array}{cc}
u^2-(1+S-iu)^2 & -2u^{2}\\
2u^{2} & -u^2+(1+S+iu)^2
\end{array}\right)\left(\begin{array}{c}
Q_{S}^{[1]}\\
Q_{S}^{[-1]}
\end{array}\right)\,,\label{eq:spinshift+} \\
S^2\left(\begin{array}{c}
Q_{S-1}^{[1]}\\
Q_{S-1}^{[-1]}
\end{array}\right) =& \left(\begin{array}{cc}
u^2-(S+iu)^2 & -2u^{2}\\
2u^{2} & -u^2+(S-iu)^2
\end{array}\right)\left(\begin{array}{c}
Q_{S}^{[1]}\\
Q_{S}^{[-1]}
\end{array}\right)\, . \label{eq:spinshift-}
\end{align}
Finally, when solving constraint equations for unknown coefficients we will need the rule for derivatives of Baxter polynomials over spectral parameter. The latter was also found empirically and has the form: 
\begin{equation}
\frac{d}{du}Q_{S}(u)=2i\sum_{k=1}^{S}(1-(-1)^{k})\left(\frac{1}{k}-\frac{1}{2S-k+1}\right)Q_{S-k}(u)\,.
\end{equation}

\subsection{Solutions of Baxter equations}

The solution of two inhomogeneous Baxter equations \eqref{eq:Baxter1} and \eqref{eq:Baxter2}, we need to solve at each perturbation order, reduce  to the solution of the following Baxter equation:
\begin{equation}
(u+\frac{i}{2})^{2}q(u+i)+(u-\frac{i}{2})^{2}q(u-i)-(2u^{2}-\frac{1}{2}-S(S+1))q(u)=V(u)\label{eq:Baxter-equation}
\end{equation}
Its first homogeneous solution  is given by (\ref{eq:Baxter-polynomial-def}).
To find the second homogeneous solution and the solution of nonhomogeneous
equation we will use the idea of dictionary \cite{ABJMtwist1}, that is we are going to  devise a recursive procedure for finding particular solutions for all types of canonical inhomogeneities, which are obtained as a result of applying elementary operations to the right hand side of Eq.\eqref{eq:Baxter-equation}. For that purpose let us introduce $B_S$ operator:
\begin{equation}
B_S[f] = (u+\frac{i}{2})^2f^{[2]} + (u-\frac{i}{2})^2f^{[-2]} - (2u^2-\frac{1}{2}-S(S+1))f\, .
\label{eq:BS-operator}
\end{equation}
Then we have
\begin{equation}
B_S \left[
Q_{S+k}
\right] = -k(2S+k+1)Q_{S+k} \label{eq:BSQ}
\end{equation}
and thus the particular solution for this type of inhomogeneity is given by 
\begin{equation}
F_S\left[Q_{S+k}\right] = B_S^{-1}\left[Q_{S+k}\right] \overset{k\neq 0}{=}-\frac{1}{k(2S+k+1)}Q_{S+k}\, . \label{eq:F-QS+k}
\end{equation}
and as a consequence for $Q$-sums\footnote{See Appendix \ref{VQandVwQsums} for a definition of $Q$ - sums.} we have
\begin{equation}
F_S\left[
\langle Q|w_1,W\rangle
\right] = \langle Q|\frac{w_1(\j)}{\j (2S-\j+1)}, W\rangle\, .
\end{equation}
To find a second homogeneous solution  consider the following relation\footnote{See Appendix \ref{Hurwitz-functions} for a definition of Hurwitz functions.}:
\begin{multline}
B_S\left[
Q_S\xi_2
\right] = 4\langle Q|1-(-1)^\j\rangle - 2\langle Q|\frac{1-(-1)^\j}{S-\j+1}\rangle + 2\langle Q|(1+2S-2\j)(1-(-1)^\j),\frac{1}{S-\j+1}\rangle
\end{multline} 
and applying $F_S$-operator as in Eq.\eqref{eq:F-QS+k} obtain
\begin{equation}
\Z_S (u) = Q_S (u)\xi_2(u) + P_S(u)\, ,
\end{equation}
where 
\begin{multline}
P_S (u) = 2\langle Q|\frac{(1-(-1)^\j)}{\j(2S-\j+1)}(2\j-2S-1)\rangle \nonumber \\
+2\langle Q|\frac{(1-(-1)^\j)}{\j (2S-\j+1)(S-\j+1)}(2\j-2S-1)\rangle
\end{multline}
Then, the general solution of homogeneous Baxter equation is given by
\begin{equation}
q^{hom}(S,u) = \Phi_1(u) Q_S(u) + \Phi_2(u) \Z_S (u)\, ,
\end{equation}
where $\Phi_a$ are arbitrary $i$-periodic functions in spectral parameter $u$. The latter are parametrized similarly to \cite{VolinPerturbativeSolution} as\footnote{See Appendix \ref{Hurwitz-functions} for definition of $\PP$ - functions.} 
\begin{equation}
\Phi_a (u) = \phi_{a,0} + \sum_{j=1}^{\Lambda_a}\phi_{a,j}\PP_j(u+i/2)\, ,
\end{equation}
where the upper summation limit depends on the order of perturbation theory and particular Baxter equation, the one for $q_1$ or $q_4$.

Next, lets see how $F_S$ images obtained for some other types of canonical inhomogeneities.    
For example, considering with the account of $u Q$ rules\footnote{See the end of Appendix \ref{VQandVwQsums}.} the relation 
\begin{equation}
B_S\left[
Q_S\xi_1
\right] = -i(1+2S)Q_S - i(1+2S)\langle Q|1+(-1)^\j\rangle 
+ 2i\langle Q|\j (1+(-1)^\j)\rangle
\end{equation}
and applying to both sides of it  $F_S$ operator we get
\begin{equation}
F_S\left[Q_S\right] = \frac{i}{2S+1}Q_S\xi_1 -\frac{1}{2s+1}\left\{
\langle Q|\frac{1+(-1)^\j}{\j}\rangle - \langle Q|\frac{1+(-1)^\j}{2S-\j+1}\rangle
\right\}\, .
\end{equation}
Similarly, considering the relation
\begin{multline}
B_S\left[
Q_S \xi_{2,a,b,A}\right] = \xi_{a,b,A}\left(Q_S^{[-2]}-Q^{[2]}\right)
+ \frac{1}{(u+i/2)^a}\xi_{b,A} \\
-\frac{1}{(u+i/2)^{a+b}}\xi_A^{[2]} + \frac{1}{(u+i/2)^a}\left(
Q_S^{[2]}-1
\right)\xi_A^{[2]}
\end{multline}
and again applying $F_S$ operator we obtain 
\begin{multline}
F_S\left[
\frac{1}{(u+i/2)^a}\xi_{b,A}
\right] = Q_S\xi_{2,a,b,A} - F_S\left[
\xi_{a,b,A}\left(Q_S^{[-2]}-Q_S^{[2]}\right)
\right] \\
+ F_S\left[
\frac{1}{(u+i/2)^{a+b}}\xi_A^{[2]}
\right] - F_S\left[
\frac{1}{(u+i/2)^a}\left(Q_S^{[2]}-1\right)\xi_{b,A}^{[2]}
\right]\, ,
\end{multline}
As a next example, let us consider $F_S$ image for $Q$ - sums multiplied by spectral parameter $u$. Using multiplication rule \eqref{eq:umult} we have
\begin{multline}
F_S\left[
u\langle Q| w_1, W\rangle 
\right] = \frac{i}{2}\sum_{j=1}^S\frac{(j-S)^2}{2S-2j+1}F_S\left[Q_{S-j-1}\right]w_1(j)|W\rangle_j \\
-\frac{i}{2}\sum_{j=1}^S\frac{(S-j+1)^2}{(2S-2j+1)}F_S\left[Q_{S-j+1}\right]w_1(j)|W\rangle_j
\end{multline}
Applying previously obtained $F_S$ images for Baxter polynomials and rewriting the result in terms of $Q^{\{n\}}$ - sums\footnote{See Appendix \ref{VQandVwQsums} for their definition.} we get
\begin{align}
F_S\left[
u\langle Q|w_1,W\rangle
\right] =& \frac{u}{(S+2)(S-1)}w_1(S)|W\rangle_S \nonumber \\
& + \frac{i}{2(2S+1)(2S+3)}\left\{
W_1^{\{-1,-1\}}\left[
\frac{1}{\j}
\right] + W_1^{\{-1,1\}}\left[
\frac{1}{2S-\j+1}
\right]
\right\} \nonumber \\
& -\frac{i}{2(2S+1)(2S-1)}\left\{
W_1^{\{-1,-1\}}\left[\frac{1}{2S-\j+1}\right]+
W_1^{\{-1,1\}}\left[\frac{1}{\j}\right]
\right\} \nonumber \\
& + \frac{u}{(2S+1)}\left\{
W_1^{\{0,1\}}\left[
\frac{1}{\j}
\right] + W_1^{\{0,1\}}\left[
\frac{1}{2S-\j+1}
\right]
\right\}\, ,
\end{align}
where $W_1^{\{n,m\}}[w] = \langle Q^{\{n\}}w(\j-m)|w_1,W\rangle_{\{1,S-1\}}$. These $Q^{\{n\}}$ - sums can be further reduced to $Q$ - sums\footnote{See Appendix \ref{VQandVwQsums} for details.}. 
Finally, consider $F_S$ image for the product of some function $f$ with $\xi_A$. In this case we start with the relation
\begin{multline}
B_S\left[\xi_A F_S\left[f\right]\right] = f\xi_A + (u+i/2)^2\left(\xi_A^{[2]}-\xi_A\right)F_S[f]^{[2]} + (u-i/2)^2\left(\xi_A^{[-2]}-\xi_A\right)F_S[f]^{[-2]}
\end{multline} 
and applying to both its sides $F_S$ operator obtain
\begin{multline}
F_S \left[
\xi_A f
\right] = \xi_A F_S [f] - F_S\left[
(u+i/2)^2(\xi_A^{[2]}-\xi_A) F_S[f]^{[2]}
\right] \\
- F_S\left[
(u-i/2)^2(\xi_A^{[-2]}-\xi_A) F_S[f]^{[-2]}
\right]
\end{multline}
The all required $F_S$ images can be be found in Appendix \ref{F-images}. Finally, we have also derived improved formula for solving this type of Baxter equations at fixed integer spins, see Appendix \ref{fixedspins} for details.

\subsection{Solution of first order difference equation}

At each perturbation order we need to solve the first order inhomogeneous difference equation of the form:
\begin{equation}
\nabla r(u) = r(u) - r(u+I) = V(u)\, .
\end{equation}
The solution of homogeneous equation is simple and  for our problem can be written as \cite{VolinPerturbativeSolution}: 
\begin{equation}
r^{\hom}(u) = \phi_{r,0} + \sum_{j=1}^{\Lambda_r}\phi_{r,j}\PP_j(u+i/2)\, ,
\end{equation}
where the upper summation limit depends on the order of perturbation theory. 

To find a particular solutions for a canonical types of inhomogeneities  obtained as a result of reduction with the help of elementary operations we need to know in first place the action of inverse to $\nabla$ operator ($\Psi$ - operator)  on Baxter polynomial. We empirically found the following rule:
\begin{equation}
\Psi\left[
Q_S
\right] = \frac{1}{2(2S+1)}\left\{
(S+1)Q_{S+1} + (2S+1)Q_S + S Q_{S-1}
\right\}\, ,
\end{equation}
Next, the straightforward application of this rule to $Q$ - sums gives 
\begin{align}
\Psi\left[
\langle Q|w_1, W\rangle
\right] =& \frac{1}{2}\langle Q|w_1,W\rangle + i u \langle Q|\frac{w_1(\j)}{S-\j+1}, W\rangle\nonumber \\ &
+ \frac{1}{2}\left\{
W_5^{\{-1,-1\}}\left[
\frac{1}{S-\j+1}
\right] - W_5^{\{-1,0\}}\left[
\frac{1}{S-\j+1}
\right]
\right\}\, ,
\end{align}
where $W_5^{\{n,m\}}[w] = \langle Q^{\{n\}}w(\j-m)|w_1, W\rangle$. The $\Psi$ - images for $Q$ - sums multiplied by spectral parameter are derived similar to $F$ - images and the results can be found in Appendix \ref{Psi-images}. Finally to obtain $\Psi$ - images for $Q$ - sums multiplied by Hurwitz functions we use the following easy to derive rule:
\begin{equation}
\Psi\left[
\xi_A\cdot h
\right] = \xi_A\Psi [h] - \Psi\left[
\left(\nabla\xi_A\right)^{[-2]}\Psi [h]
\right]^{[2]}\, , \label{eq:Psi-h-xi-relation}
\end{equation}
where $h$ is some polynomial build from Baxter polynomials and their sums. Indeed, considering the relation 
\begin{equation}
\nabla \left(\xi_A\Psi[h])\right) = \xi_A\Psi [h] - \xi^{[2]}\Psi\left[
h^{[2]}
\right] = \xi_A h + \nabla\xi_A\Psi\left[h^{[2]}\right]
\end{equation}
and applying to both its sides $\Psi$ operator we get Eq.\eqref{eq:Psi-h-xi-relation}. The other required rules together with those already derived in \cite{VolinPerturbativeSolution} and can be found in Appendix \ref{Psi-images}.

\section{Constraints solution and anomalous dimensions}
\label{constraintsSolutionAnomDimensions}

Now, that we know how to solve inhomogeneous Baxter and first order difference equations the recursive procedure for the solution of $\P\mu$ - system proceeds through the following steps at each order of perturbation theory\footnote{They are similar to those presented in \cite{VolinPerturbativeSolution}.}:
\begin{enumerate}
\item Solve first Baxter equation for $q_1$ \eqref{eq:Baxter1} and determine all coefficients $\phi_{1,k}^{q_1,(l)}$ and $\phi_{2,k}^{q_1,(l)}$ except $\phi_{1,0}^{q_1,(l)}$ from consistency relations for the absence of poles at $u=0$ for a combination of functions \eqref{eq:mupoles1}-\eqref{eq:mupoles2} written for $\mu_1$. The coefficient $\phi_{1,0}^{q_1,(l)}$ is then determined from the constraint for $\Pt_1$ - function \eqref{eq:Pt1} expanded at $u=0$ up to $\OO (u^3)$.
\item Solve first order difference equation for $r_2$ \eqref{eq:firstorder} and determine the coefficients $\phi_{r,k}^{(l)}$ from the constraint for $\Pt_2$ - function expanded at $u=0$ up to $\OO (1)$. The corresponding constraint follows from the $r_2$ solution and ansatz for $\P_2$ function.
\item Determine coefficients $A_4^{(l)}$ and $n_1^{(l)}$ from the constraint for $\Pt_2$ - function expanded at $u=0$ up to $\OO (u^3)$. The knowledge of $A_4^{(l)}$ coefficient then allows us to determine anomalous dimension $\gamma^{(l+1)}$ from Eq. \eqref{eq:A3A4}.
\item Solve second Baxter equation for $q_4$ \eqref{eq:Baxter2} and determine all coefficients $\phi_{1,k}^{q_4,(l)}$ and $\phi_{2,k}^{q_4,(l)}$ except $\phi_{1,0}^{q_4,(l)}$ from consistency relations for the absence of poles at $u=0$ for a combination of functions \eqref{eq:mupoles1}-\eqref{eq:mupoles2} written for $\mu_4$. The coefficient $\phi_{1,0}^{q_4,(l)}$ is then determined from constraint for $\Pt_4$ - function \eqref{eq:Pt4} expanded at $u=0$ up to $\OO (u^3)$.
\item Determine coefficients $c_{1,k}^{(l)}$, $c_{2,k}^{(l)}$, $c_{3,k}^{(l)}$, $c_{4,k}^{(l)}$
from constraints for $\Pt_1$, $\Pt_2$, $\Pt_3$, $\Pt_4$ - functions expanded at $u=0$ up to required order.
\item Proceed with the next iteration. 
\end{enumerate}
Here, the iteration order $l$ corresponds to loop order $l+1$ and we start iterative procedure with $l=0$. Note also, that performing series expansions of various constraints one should also take into account relations between multiple zeta functions \cite{zetaRelations}.

Following the procedure described we obtained the following known expressions for anomalous dimensions of twist 2 operators up to four loop order \cite{N4SYM1loop,N4SYM2loop1,N4SYM2loop2,N4SYM2loop3,N4SYM2loop4,N4SYM3loop,N4SYM4loop}: 
\begin{equation}
\gamma (S) = \PP\left(S+\frac{1}{2}\gamma (S)\right)\, ,
\end{equation}
where
\begin{align}
\gamma (S) = \sum_{l=1}^{\infty}g^{2l}\gamma^{(l)}(S)\, , \qquad
\PP (S) = \sum_{l=1}^{\infty} g^{2l} \PP^{(l)} (S)\,
\end{align}
and reciprocity respecting \cite{GLreciprocity1,GLreciprocity2,GLreciprocity3,GLreciprocity4,GLreciprocity5,GLreciprocity6,GLreciprocity7} function $\PP(S)$ up to four loop order is given by
\begin{align}
\frac{\PP^{(1)}(S)}{4} =& \H_1\, , \\
\frac{\PP^{(2)}(S)}{8} =& \H_{1,2} - \H_{2,1}\, , \\
\frac{\PP^{(3)}(S)}{16} =& -\H_{2,3} + \H_{4,1} + 2\H_{1,1,3} -3\H_{1,2,2} + \H_{1,3,1} -2\H_{2,1,2} + \H_{2,2,1} + \H_{3,1,1} \\
\frac{\PP^{(4)}(S)}{32} =& -2\H_{1,2,4} +2\H_{1,5,1} + 4\H_{2,2,3} + \H_{2,3,2} - 5\H_{2,4,1}
+2\H_{3,1,3} - 3H_{3,2,2} \nonumber \\ &
-\H_{3,3,1} +4\H_{4,1,2} -4\H_{4,2,1} +2\H_{5,1,1} +4\H_{1,1,1,4} -6\H_{1,1,2,3} \nonumber \\&
-2\H_{1,1,3,2} +4\H_{1,1,4,1} -2\H_{1,2,1,3} +9\H_{1,2,2,2} -9\H_{1,2,3,1} +3\H_{1,3,1,2} \nonumber \\ &
-5H_{1,3,2,1} +4\H_{1,4,1,1} -2\H_{2,1,1,3} +5\H_{2,1,2,2} -5\H_{2,1,3,1} +8\H_{2,2,2,1}\nonumber \\ & -6\H_{2,3,1,1} +3\H_{3,1,1,2} -\H_{3,1,2,1} -4\H_{3,2,1,1} +2\H_{4,1,1,1} -4\H_{1,1,1,2,2} \nonumber \\ &
+ 6\H_{1,1,1,3,1} + 2\H_{1,1,2,1,2} -8\H_{1,1,2,2,1} +4\H_{1,1,3,1,1} + 2\H_{1,2,1,1,2}\nonumber \\ &
-4\H_{1,2,2,1,1} +2\H_{1,3,1,1,1} + 2\H_{2,1,1,1,2} -2\H_{2,2,1,1,1} -10\H_{1,1}\zeta_5\nonumber \\ &
+ (8\H_4 + 6\H_{1,3} -6\H_{2,2} + 4\H_{1,1,2} -2\H_{1,2,1} -2\H_{2,1,1})\zeta_3\, .
\end{align}
Here $\H_{i_1,\ldots , i_k}$ stand for binomial harmonic sums defined as \cite{Vermaseren}
\begin{equation}
\H_{i_1,\ldots ,i_k}\equiv \H_{i_1,\ldots ,i_k}(S) = (-1)^S\sum_{j=1}^S(-1)^j\binom{S}{j}\binom{S+j}{j}H_{i_1,\ldots ,i_k}(S)\, \\
\end{equation}
and $H_{i_1,\ldots ,i_k}(S)$ are usual nested harmonic sums \cite{Vermaseren}.

\section{Conclusion}\label{Conclusion}

So, let us summarize the obtained results.  We introduced the class of functions - products of rational functions in spectral parameter $u$ with sums of Baxter polynomials and Hurwitz functions closed under elementary operations, such as argument shifts and partial fractions, as well as under differentiation. Next, we have shown that this class of functions is sufficient for the perturbative solution of Riemann-Hilbert problem for $\P\mu$ - system for twist 2 operators in $sl(2)$ sector in the case of $\N=4$ SYM model and arbitrary integer spin values. In particular we have presented a recursive procedure based on the idea of dictionary \cite{ABJMtwist1} for the perturbative solution of arising inhomogeneous Baxter and first order difference equations. In the present paper we have limited ourselves by the calculation of anomalous dimensions up to four loop order only. To go further and in particular beyond already available 5, 6 and 7 loop results \cite{N4SYM5loop,N4SYM5loop2,N4SYM6loop,N4SYM7loop} an optimization of our {\it Mathematica} code and its realization for example in language of FORM computer algebra system \cite{FORM} would be desirable. 
 
Our calculation strategy is a generalization of fixed spin approach of \cite{VolinPerturbativeSolution} to the case of arbitrary integer spin values, where spin is treated as arbitrary parameter. On the other hand there is an alternative fixed spin approach of \cite{N4QSCnumeric1,GromovBFKL2,BFKLnonzeroConformalSpin}. In the latter case one considers a system of first order difference equations for $Q_{a|i}$ - functions ($a,i$=1,\ldots,4): 
\begin{equation}
Q_{a|i}(u-i/2) = (\delta_a^b - \P_a(u)\P_c(u)\chi^{cb})Q_{b|i}(u+i/2)\, . \label{eq:Qai-system}
\end{equation} 
The $\P_a$ - functions are again parametrized as in \eqref{eq:Pansatz1}-\eqref{eq:Pansatz4}. After solving this system of equations the undetermined coefficients in the parametrization of $\P$ - functions are obtained from the gluing conditions for $\Q_i$ and $\Qt_i$ - functions ($i=1,\ldots ,4$):
\begin{align}
\Q_i(u) =& Q_{a|i}(u+i/2)\chi^{ab}\P_b (u)\, ,\\ \Qt_i (u) =& Q_{a|i}(u+i/2)\chi^{ab}\Pt_b (u)\, .
\end{align}
To solve perturbatively the system \eqref{eq:Qai-system} \cite{GromovBFKL2,BFKLnonzeroConformalSpin} uses variation of constants method and search for the solution with the ansatz
\begin{equation}
Q_{a|i} = Q_{a|i}^{(0)} + b_i^{j}(u+i/2)Q_{a|j}^{(0)}\, ,
\end{equation}
where $Q_{a|i}^{(0)}$ is the zero order solution for $Q_{a|i}$ - function given by the solution of corresponding forth-order Baxter equation. Then to find $b_i^j$ we will need to solve only first order difference equations. At fixed integer spins the solution for $b_i^j$ can be obtained in terms of rational functions in spectral parameter times Hurwitz functions. To obtain the solution for $Q_{a|i}$ we still need to multiply this solution by $Q^{(0)}_{a|i}$. At fixed integer values this operation leaves us within the same class of functions due to stuffle relations for Hurwitz functions. At arbitrary spin values on the other hand we will need to multiply sums of $Q_{a|i}^{(0)}$ - functions and for example for sums of Baxter polynomials considered in this paper we failed to find such kind of relations, which would help us to stay within originally specified class of functions. So, it seems that any calculation for arbitrary spin values of twist 2 operators will require consideration of at least second order inhomogeneous  difference equations and their solution.  

We expect the presented method to be generalizable to higher twists as well as to other theories, such as ABJM model. In the later case for twist 2 operators we were able to find all required elementary operations for corresponding Baxter polynomials, except differentiation with respect to spectral parameter. It was not also possible to find a finite set of rules for required $F_1$ and $F_2$ images, see \cite{uspaceTMF,ABJMtwist1} for their definition. We will present our finding in one of our future publications.  Moreover, similar ideas should be also applicable to the study of BFKL regime within quantum spectral curve approach \cite{GromovBFKL1,GromovBFKL2,BFKLnonzeroConformalSpin}. In this case we also have a perturbative expansion when both coupling constant $g$ and parameter $w\equiv S+1$, describing the proximity of operator spin $S$ to $-1$ are considered to be small, while  their ratio $g^2/w$ remains fixed.

\section*{Acknowledgements}

I would like to thank Roman Lee for interest to this work and collaboration on similar problems within context of ABJM model and twist 1 operators, which was a big step towards present solution. This work was supported by Foundation for the Advancement of Theoretical Physics and Mathematics "BASIS". 

\appendix

\section{Hurwitz functions}\label{Hurwitz-functions}

 We define Hurwitz functions entering the presented solution as\footnote{All indexes in the case of $\N=4$ SYM QSC are positive.} 
 \begin{align}\label{eq:xiseries}
 \xi_{a,A} & =\sum_{n=1}^{\infty}\frac{1}{\left(u+in-\frac{i}{2}\right)^{a}}\xi_{A}^{\left[2n\right]}.
 \end{align}
Here $A$ denotes the arbitrary sequence of indexes and $\xi$ function without indexes is identical to unity.  These are the shifted versions of Hurwitz functions introduced in \cite{VolinPerturbativeSolution,ABJMQSC12loops}
 \begin{equation}
 \xi_{A}=\eta_{A}^{\left[1\right]}
 \end{equation}
 The $\xi_{1\ldots 1}$ functions should be defined separately, as the series \eqref{eq:xiseries} diverge in this case. For $\xi_1$ - function we have 
 \begin{equation}
 \xi_{1}\left(u\right)=i\psi\left(-iu+\tfrac{i}{2}\right)\,.
 \end{equation}
and $\xi_{1\ldots 1}$ functions are defined as \cite{VolinMZVdoublewrapping}:
\begin{equation}
\xi_{\underbrace{1,\ldots,1}_{k}}\left(u\right)=\frac{1}{k!}\left(\xi_{1}+\partial_{u}\right)^{k}1\,.
\end{equation} 	
For shifts of our Hurwitz functions we have
\begin{align}\label{eq:xishift}
\xi_{a,A}^{\left[2\right]} & =\xi_{a,A}-\frac{1}{\left(u+\frac{i}{2}\right)^{\left|a\right|}}\xi_{A}^{\left[2\right]}\\
\xi_{a,A}^{\left[-2\right]} & =\xi_{a,A}+\frac{1}{\left(u-\frac{i}{2}\right)^{\left|a\right|}}\xi_{A}.
\end{align} 
We will also need the following $i$-periodic combinations of Hurwitz functions $\xi_k$ defined as 
\begin{equation}
\PP_k (u) = \xi_k^{[-1]}(u) + (-)^k \xi_k^{[-3]}(-u) = \PP_k (u+i)\, , \quad k > 0 \in \mathbb{Z}\, ,
\end{equation}
Note that $\PP_k (u)$ can be expressed via elementary functions:
\begin{equation}
\PP_k (u) = \frac{(-\partial_u)^{k-1}}{(k-1)!} 
\pi\coth(\pi u)
\end{equation} 

\section{$Q$ and $Q^{\{n\}}$ - sums}\label{VQandVwQsums}

The whole perturbative solution of Riemann-Hilbert problem for $\P\mu$ - system considered in present paper was performed in a special class of functions, which includes sums of Baxter polynomials.  First, we define $Q$ - sums similar to \cite{uspaceTMF,ABJMtwist1} as
\begin{multline}
\left\langle Q\left(u\right)|w_{1}\left(\j\right),\ldots,w_{n}\left(\j\right)\right\rangle_{\{S_Q,S_d,S_{u}\}}   = \sum_{j1=S_d}^{S_u} Q_{S_Q-j_1}\left(u\right)
\sum_{j_{1}>j_{2}\ldots>j_{n}>0}\prod_{k}w_{k}\left(j_{k}\right) \\
 \sum_{j_1=S_d}^{S_u}Q_{S_Q-j_1}\left(u\right) w_1(j_1)|w_2(\j),\ldots ,w_n(\j)\rangle_{j_1}\,  
 , \label{QW-sums} 
\end{multline} 
\begin{gather} 
\left\langle Q\left(u\right)|w_{1}\left(\j\right),\ldots,w_{n}\left(\j\right)\right\rangle = 
\left\langle Q\left(u\right)|w_{1}\left(\j\right),\ldots,w_{n}\left(\j\right)\right\rangle_{\{S,1,S\}} \\
\left\langle Q\left(u\right)|\right\rangle = Q_{S}\left(u\right), 
\end{gather}
where $w_k$ are some weights. The bullet $\j$ denotes summation index or argument of weight function $w_k$, whose particular symbolic representation or name is not important. Here and below we write weights $w_k(\j)$ in several equivalent ways, $w_k (\j) \equiv w_k (j) \equiv w_k$ and use $W$ to denote arbitrary (maybe empty) sequence of weights. For example, we have
\begin{equation}
\left\langle Q\left(u\right)|\frac{(-1)^\j}{(\j)^3}, \frac{1}{(S+1-\j)^2} \right\rangle = \sum_{S\geq j_{1}>j_{2}>0}Q_{S-j_1}(u) \frac{(-1)^{j_1}}{j_1^3}\frac{1}{(S+1-j_2)^2}
\end{equation}
Also above we implicitly used the notation
\begin{equation}\label{eq:ketsums}
\left|w_{1},w_{2},\ldots,w_{n}\right\rangle_{j_0} =\sum_{j_{0}>j_{1}\ldots>j_{n}>0}\prod_{k}w_{k}(j_k)\,,
\end{equation}
In the case, when the argument of $Q_S$ is $u$ we will often drop it and simply write $\left\langle Q|w_{1}\left(\j\right),w_{2}\left(\j\right),\ldots,w_{n}\left(\j\right)\right\rangle $.
We also introduce a shortcut
\begin{align}
\left\langle w_{1}\left(\j\right),w_{2}\left(\j\right),\ldots,w_{n}\left(\j\right)\right\rangle  & =\left\langle Q(\tfrac{i}{2})| w_{1}\left(\j\right),w_{2}\left(\j\right),\ldots,w_{n}\left(\j\right)\right\rangle\,
=\sum_{S\geq j_{1}>j_{2}\ldots>j_{n}>0}\prod_{k}w_{k}\,.
\end{align} 
Note that the  $\left\langle w_{1},W\right\rangle$-sums satisfy usual stuffle relations.

It turns out that the weights in sums at twist 2 for $\N=4$ SYM model can be always reduced to the same set of weights we had previously at twist 1 for ABJM model \cite{uspaceTMF,ABJMtwist1}. The latter are given by
\begin{gather}\label{eq:cweights}
\frac{1}{\j^{n}}=n_{+}(\j)\,,\qquad\frac{\left(-\right)^{\j}}{\j^{n}}=n_{-}(\j)\,,\\
\frac{1}{\left(S+1-\j\right)^{n}}=\overline{n}_{+}(\j)\,,\qquad\frac{\left(-\right)^{\j}}{\left(S+1-\j\right)^{n}}=\overline{n}_{-}(\j)\,,\\
\frac{1}{\left(2S+1-\j\right)^{n}}=\hat{n}_{+}(\j)\,,\qquad\frac{\left(-\right)^{\j}}{\left(2S+1-\j\right)^{n}}=\hat{n}_{-}(\j)\,.
\end{gather}
Next, we will need the so call $Q^{\{n\}}$ - sums. The latter are defined as sums of $Q^{\{n\}}$ polynomials
\begin{equation}
Q_{S}^{\{n\}}=\left(\frac{\Gamma(S+n+1)}{\Gamma(S+1)}\right)^{2\,\sign(n)}Q_{S+n}\, ,
\end{equation}
which have the following properties:
\begin{align}
Q_{S+1}^{\{n\}} & =(S+1)^{2}Q_{S}^{\{n+1\}}\,,\;n<0 \label{eq:Qnshift1}\\
Q_{S-1}^{\{n\}} & =\frac{1}{S^{2}}Q_{S}^{\{n-1\}}\,,\;n<0 \label{eq:Qnshift2}\\
Q_{S-1}^{\{n\}} & =S^{2}Q_{S}^{\{n-1\}}\,,\;n>0 \label{eq:Qnshift3}\\
Q_{S+1}^{\{n\}} & =\frac{1}{(S+1)^{2}}Q_{S}^{\{n+1\}}\,,\;n>0 \label{eq:Qnshift4}
\end{align}
and
\begin{align}
Q_{S}^{\{n\}} & =(S+n-1)^{4}Q_{S}^{\{n-2\}}+2i(2S+2n-1)uQ_{S}^{\{n-1\}}\,,\;n>1 \label{eq:nshift1}\\
Q_{S}^{\{1\}} & =\frac{1}{2}(2u-1-(1-i)S)(2u+1+(1+i)S)Q_{S}-\frac{1}{2}(2u-i)^{2}Q_{S}^{[-2]}\,,
\label{eq:nshift2}\\
Q_{S}^{\{0\}} & =Q_{S}\,,
\label{eq:nshift3}\\
Q_{S}^{\{-1\}} & =\frac{i}{2}(i+(1+i)S-2u)(1+(1+i)S+2iu)Q_{S}-\frac{1}{2}(2u-i)^{2}Q_{S}^{[-2]}\,,
\label{eq:nshift4}\\
Q_{S}^{\{n\}} & =-2i(3+2n+2S)uQ_{S}^{\{n+1\}}+(S+n+2)^{4}Q_{S}^{\{n+2\}}\,.
\label{eq:nshift5}
\end{align}
For $Q^{\{n\}}$ - sums themselves we will use the following notation:
\begin{align}
\langle Q^{\{n\}}w_{0}(\j-m)|w_{1},W\rangle_{S_d,S_u} & =\sum_{j=S_d}^{S_u}Q_{S-j}^{\{n\}}w_{0}(j-m)w_{1}(j)|W\rangle_{j}\, , \\
\langle Q^{\{n\}}w_{0}(\j-m)|w_{1},W\rangle & = \langle Q^{\{n\}}w_{0}(\j-m)|w_{1},W\rangle_{1,S}\,,\;m>0\\
\langle Q^{\{n\}}w_{0}(\j-m)|w_{1},W\rangle & = \langle Q^{\{n\}}w_{0}(\j-m)|w_{1},W\rangle_{1,S+m}\,,\;m<0
\end{align}
These new sums can be further reduced to previously defined $Q$ - sums. This is done via the following steps. First, we reduce their summation limits with
\begin{align}
\langle Q^{\{n\}}w_0(\j-m)|w_1,W\rangle_{\{S_d,S_u\}} \overset{m\geq 0, S_u<S}{=} &
\langle Q^{\{n\}}w_0 (\j-m)|w_1,W\rangle_{\{S_d, S\}} \nonumber \\ & 
-\sum_{k_{S_u}+1}^S Q_{S-k}^{\{n\}} w_0 (k-m) w_1(k) |W\rangle_k\, ,\label{eq:VwQlim1} \\
\langle Q^{\{n\}}w_0(\j-m)|w_1,W\rangle_{\{S_d,S_u\}} \overset{m < 0, S_u < S+m}{=} & \langle Q^{\{n\}}w_0(\j-m)|w_1,W\rangle_{\{S_d,S+m\}} \nonumber \\ &
-\sum_{k_{S_u}+1}^{S+m} Q_{S-k}^{\{n\}} w_0 (k-m) w_1(k) |W\rangle_k\, , \label{eq:VwQlim2} \\ 
Q^{\{n\}}w_0(\j-m)|w_1,W\rangle_{\{S_d,S\}} \overset{m\geq 0, S_d > 1}{=} &
\langle Q^{\{n\}}w_0(\j-m)|w_1,W\rangle \nonumber \\ &
-\sum_{k=1}^{S_d-1} Q_{S-k}^{\{n\}} w_0 (k-m) w_1(k)|W\rangle_k \, , \label{eq:VwQlim3} \\
Q^{\{n\}}w_0(\j-m)|w_1,W\rangle_{\{S_d,S+m\}} \overset{m < 0, S_d > 1}{=} &
\langle Q^{\{n\}}w_0(\j-m)|w_1,W\rangle \nonumber \\ &
-\sum_{k=1}^{S_d-1} Q_{S-k}^{\{n\}} w_0 (k-m) w_1(k)|W\rangle_k \, , \label{eq:VwQlim4}
\end{align}
In a third rule it is assumed that $w_0(j) \neq \frac{1}{j}$ or $\frac{(-1)^j}{j}$. If $w_0(j) = \frac{1}{j}$ we use instead
\begin{align}
\langle Q^{\{n\}}\frac{1}{\j-m}|w_1,W\rangle_{S_d,S} \overset{m\geq 0, S_d > m+1}{=} & \langle Q^{\{n\}}\frac{1}{\j-m}|w_1,W\rangle_{\{m+1,S\}} \nonumber \\ &
-\sum_{k=m+1}^{S_d-1}Q_{S-k}^{\{n\}}\frac{1}{k-m}w_1(k)|W\rangle_k\, , \label{eq:VwQlim5} \\
\langle Q^{\{n\}}\frac{1}{\j-m}|w_1,W\rangle_{\{m+1,S\}} \overset{m>0}{=} & 
w_1(m)\langle Q^{\{n\}}\frac{1}{\j-m}|1,W\rangle \nonumber \\ &
+ \langle Q^{\{n\}}|\frac{w_1(\j)-w_1(m)}{\j-m},W\rangle_{\{m+1,S\}}\, , \label{eq:VwQlim6} \\
\langle Q^{\{n\}}\frac{1}{\j-m}|1,w_2,W\rangle_{\{m+1,S\}} \overset{m>0}{=} & \frac{1}{(S+1-m)^2}\langle Q^{\{n-1\}}\frac{1}{\j-m+1}|1,w_2,W\rangle_{\{m,S\}} \nonumber \\ &
+ \frac{1}{(S+1-m)^2}\langle Q^{\{n-1\}}\frac{1}{\j-m+1}|w_2,W\rangle_{\{m,S\}} \nonumber \\ &
+ \frac{1}{(S+1-m)^2}\langle Q^{\{n\}}|(2S+2-\j-m),w_2,W\rangle_{\{m+1,S\}} \nonumber \\ &
-\frac{1}{(S+1-m)^3} Q_{n-1}(1+w_2(S))|W\rangle_S\, , \label{eq:VwQlim7} \\
\langle Q^{\{n\}}\frac{1}{\j-m}|1,w_2\rangle_{\{m+1,S\}} \overset{m>0}{=} & 
\frac{1}{(S+1-m)^2}\langle Q^{\{n-1\}}\frac{1}{\j-m+1}|1,w_2\rangle_{\{m,S\}} \nonumber \\ &
+ \frac{1}{(S+1-m)^2}\langle Q^{\{n-1\}}\frac{1}{\j-m+1}|w_2\rangle_{\{m,S\}} \nonumber \\ &
+ \frac{1}{(S+1-m)^2}\langle Q^{\{n\}}|(2S+2-\j-m),w_2\rangle_{\{m+1,S\}} \nonumber \\ &
-\frac{1}{(S+1-m)^3}Q_{n-1} |W\rangle_{S+1}\, , \label{eq:VwQlim8} \\
\langle Q^{\{n\}}\frac{1}{\j-m}|1\rangle_{\{m+1,S\}} \overset{m>0}{=} & 
\frac{1}{(S+1-m)^2}\langle Q^{\{n-1\}}\frac{1}{\j-m+1}|1\rangle_{\{m,S\}} \nonumber \\ &
+ \frac{1}{(S+1-m)^2}\langle Q^{\{n\}}|(2S+2-\j-m)\rangle_{\{m+1,S\}} \nonumber \\ & - \frac{1}{(S+1-m)^3}Q_{n-1}\, . \label{eq:VwQlim9}
\end{align}
and similar rules when $w_0(j)=\frac{(-1)^j}{j}$.

At next step we reduce all weights in $Q^{\{n\}}$ - sums to canonical weights (reduce $m$ to zero) with the relations\footnote{$w_0$ weights in $Q^{\{n\}}$ - sums we will encounter have at most simple poles.} ($j_0$ is the pole of $w_0(\j-m)=w_0(j-m)$):
\begin{align}
\langle Q^{\{n\}}w_0(\j-m)|w_1,W\rangle \overset{m>0}{=} & 
\langle Q^{\{n\}} w_0(\j-m)w_1(j_0)|1,W\rangle \nonumber \\ &
+ \langle Q^{\{n\}}|w_0(\j-m)(w_1(\j)-w_1(j_0)), W\rangle\, , \label{eq:VwQtocanon1} \\
\langle Q^{\{n\}}w_0(\j-m)|w_1,W\rangle \overset{m<0}{=} & 
\langle Q^{\{n\}}w_0(\j-m)w_1(j_0)|1,W\rangle \nonumber \\ &
+ \langle Q^{\{n\}}|w_0(\j-m)(w_1(\j)-w_1(j_0)),W\rangle \nonumber \\ &
-\sum_{k=S+m+1}^S Q_{S-k}^{\{n\}}(w_1(k)-w_1(j_0))|W\rangle_k\, ,  \label{eq:VwQtocanon2}
\end{align}
and ($w_1(j)$ is polynomial in $j$ and $\wt0(j)=w_0(S+1-m)+w_0^{'}(S+1-m)(j-S)$)
\begin{align}
\langle Q^{\{n\}}w_0(\j-m)|w_1,w_2,W\rangle \overset{n\leq 0, m>0}{=} & 
\langle Q^{\{n\}}|\wt0 (\j-1) w_1 (\j),w_2,W\rangle \nonumber \\ &
+ \langle Q^{\{n-1\}}\frac{w_0(\j-m+1)-\wt0(\j)}{(S-\j)^2}|w_1(\j+1),w_2,W\rangle \nonumber \\ &
+ \langle Q^{\{n-1\}}\frac{w_0(\j-m+1)-\wt0(\j)}{(S-\j)^2}|w_1(\j+1)w_2(\j),W\rangle\, , \label{eq:VwQtocanon3} \\
\langle Q^{\{n\}}w_0(\j-m)|w_1\rangle  \overset{n\leq 0, m>0}{=} &  
\langle Q^{\{n\}}|\wt0 (\j-1) w_1 (\j)\rangle \nonumber \\ &
+ \langle Q^{\{n-1\}}\frac{w_0(\j-m+1)-\wt0(\j)}{(S-\j)^2}|w_1(\j+1)\rangle \nonumber \\ &
+ Q_{S+n-1}\frac{\left((S+1)_{1-n}\right)^2}{S^2}(w_0(1-m)-\wt0(0))w_1(1) \, ,
\label{eq:VwQtocanon4} \\
\langle Q^{\{n\}} w_0(\j-m)|w_1,w_2,W\rangle \overset{n<0, m<0}{=} & \langle Q^{\{n+1\}} w_0 (\j-m-1)|(S-\j+1)^2w_1(\j-1),w_2,W\rangle \nonumber \\ &
- \langle Q^{\{n\}}w_0 (\j-m)| w_1\cdot w_2, W\rangle \, ,
\label{eq:VwQtocanon5} \\
\langle Q^{\{n\}} w_0(\j-m)|w_1\rangle \overset{n<0, m<0}{=} & 
\langle Q^{\{n+1\}} w_0 (\j-m-1)|(S-\j+1)^2w_1(\j-1)\rangle \nonumber \\ &
-Q_{S+n} S^2 \left((1+n+S)_{-n-1}\right)^2 w_1(0)w_0(-m) \, ,
\label{eq:VwQtocanon6} \\
\langle Q^{\{n\}} w_0 (\j-m)|w_1,w_2,W\rangle \overset{n>0, m>0}{=} & 
\langle Q^{\{n-1\}}w_0(\j-m+1)|(S-\j)^2w_1(\j+1),w_2,W\rangle \nonumber \\ &
+ \langle Q^{\{n-1\}}w_0(\j-m+1)|(S-\j)^2w_1(\j+1)w_2(\j),W\rangle \, ,
\label{eq:VwQtocanon7} \\
\langle Q^{\{n\}} w_0 (\j-m)|w_1\rangle \overset{n>0, m>0}{=} & 
\langle Q^{\{n-1\}}w_0(\j-m+1)|(S-\j)^2w_1(\j+1)\rangle \nonumber \\ &
+ Q_{S+n-1} S^2 \left((1+S)_{n-1}\right)^2 w_0(1-m)w_1(1) \, ,
\label{eq:VwQtocanon8} \\
\langle Q^{\{n\}} w_0 (\j-m)|w_1,w_2,W\rangle \overset{n\geq 0, m<0}{=} & 
\langle Q^{\{n+1\}}w_0(\j-m-1)|\frac{w_1(\j-1)}{(S-\j+1)^2},w_2,W\rangle \nonumber \\ &
-\langle Q^{\{n\}}w_0 (\j-m)|w_1\cdot w_2, W\rangle \, ,
\label{eq:VwQtocanon9} \\
\langle Q^{\{n\}} w_0 (\j-m)|w_1\rangle \overset{n\geq 0, m<0}{=} & 
\langle Q^{\{n+1\}}w_0(\j-m-1)|\frac{w_1(\j-1)}{(S-\j+1)^2}\rangle \nonumber \\ &
- Q_{S+n}\frac{\left((S)_{n+1}\right)^2}{S^2}w_0(-m)w_1(0)\, .
\label{eq:VwQtocanon10}
\end{align}
When all weights reduced to canonical we have
\begin{equation}
\langle Q^{\{n\}}w_0(\j)|w_1, W\rangle = \langle Q^{\{n\}}|w_0(\j)w_1(\j), W\rangle \, .
\end{equation}
Finally, the $n$ index of $Q^{\{n\}}$ in $Q^{\{n\}}$ - sums is reduced with  the simple sequence of formula \eqref{eq:nshift1}-\eqref{eq:nshift5} :
\begin{align}
\langle Q^{\{n\}}|w_{1},W\rangle & \overset{n>1}{=}\langle Q^{\{n-2\}}|(S+n-\j-1)^{4}w_{1}(\j),W\rangle\nonumber \\
& +2iu\langle Q^{\{n-1\}}|(2S-2\j+2n-1)w_{1}(\j),W\rangle\,,\label{eq:VwQindex1}\\
\langle Q^{\{n\}}|w_{1},W\rangle & \overset{n<-1}{=}\langle Q^{\{n+2\}}|(S+n-\j-1)^{4}w_{1}(\j),W\rangle\nonumber \\
& -2iu\langle Q^{\{n+1\}}|(2S-2\j+2n+3)w_{1}(\j),W\rangle\,,\label{eq:VwQindex2}\\
\langle Q^{\{1\}}|w_{1},W\rangle & =(-\frac{1}{2}-S-S^{2}+2iSu+2u^{2})\langle Q|w_{1},W\rangle\nonumber \\
& +(\frac{1}{2}+2iu-2u^{2})\langle Q|w_{1},W\rangle^{[-2]}\nonumber \\
& +(1+2S-2iu)\langle Q|\j w_{1}(\j),W\rangle-\langle Q|\j^{2}w_{1}(\j),W\rangle\,,\label{eq:VwQindex3}\\
\langle Q^{\{-1\}}|w_{1},W\rangle & =(-\frac{1}{2}-S-S^{2}-2iu-2iSu+2u^{2})\langle Q|w_{1},W\rangle\nonumber \\
& +(\frac{1}{2}+2iu-2u^{2})\langle Q|w_{1},W\rangle^{[-2]}\nonumber \\
& +(1+2S+2iu)\langle Q|\j w_{1}(\j),W\rangle-\langle Q|\j^{2} w_{1}(\j),W\rangle\,,
\label{eq:VwQindex4}
\end{align}
When $w_1(j)$ is polynomial in $j$ we may also use the following reduction formula:
\begin{align}
\langle Q^{\{n\}}|w_{1},w_{2},W\rangle & \overset{n>0}{=}\langle Q^{\{n-1\}}|(S-\j)^{2}w_{1}(\j+1)w_{2}(\j),W\rangle \nonumber \\
& +\langle Q^{\{n-1\}}|(S-\j)^{2}w_{1}(\j+1),w_{2},W\rangle\,, \label{eq:VwQindex5}\\
\langle Q^{\{n\}}|w_{1}\rangle & \overset{n>0}{=}\langle Q^{\{n-1\}}|(S-\j)^{2}w_{1}(\j+1)\rangle\nonumber \\ & +Q_{S+n-1}\left((1+S)_{n-1}\right)^{2}S^{2}w_{1}(1)\,,
\label{eq:VwQindex6}\\
\langle Q^{\{n\}}|w_{1},w_{2},W\rangle & \overset{n<0}{=}\langle Q^{\{n+1\}}|(S-\j+1)^{2}w_{1}(\j-1),w_{2},W\rangle -\langle Q^{\{n\}}|w_{1}w_{2},W\rangle\,,
\label{eq:VwQindex7}\\
\langle Q^{\{n\}}|w_{1}\rangle & \overset{n<0}{=}\langle Q^{\{n+1\}}|(S-\j+1)^{2}w_{1}(\j-1)\rangle\nonumber \\ &-Q_{S+n}\left((1+n+S)_{-n-1}\right)^{2}S^{2}w_{1}(0)\,. \label{eq:VwQindex8}
\end{align} 
To understand how reduction rules for $Q^{\{n\}}$ - sums were obtained lets consider as example derivation of Eq. \eqref{eq:VwQtocanon9}. First we perform shift of first summation index as
\begin{multline}
\langle Q^{\{n\}}w_0(j-m)|w_1,w_2,W\rangle \overset{n>0, m<0}{=}
\sum_{j=2}^{S+1}Q_{S-j+1}^{\{n\}}w_0 (j-m-1)w_1(j-1)|w_2,W\rangle_{j-1} = \\
\sum_{j=2}^{S+1}Q_{S-j+1}^{\{n\}}w_0(j-m-1)w_1(j-1)|w_2,W\rangle_j - \sum_{j=2}^{S+1}Q_{S-j+1}^{\{n\}}w_0(j-m-1)w_1(j-1)w_2(j-1)|W\rangle_{j-1}
\end{multline}  
Next, in the first sum we shift index $n$ of $Q^{\{n\}}$ function with the help of Eq.\eqref{eq:Qnshift4}, while in the second sum shift summation index back and obtain
\begin{multline}
= \sum_{j=2}^{S+1}Q_{S-j}^{\{n+1\}}\frac{w_0(j-m-1)w_1(j-1)}{(S-j+1)^2}|w_2,W\rangle_j
- \sum_{j=1}^S Q_{S-j}^{\{n\}}w_0(j-m)w_1(j)w_2(j)|W\rangle_j \\
= \langle Q^{\{n+1\}}w_0(j-m)|\frac{w_1(j-1)}{(S-j+1)^2},w_2,W\rangle - \langle Q^{\{n\}}w_0(j-m)|w_1\cdot w_2, W\rangle\, .
\end{multline}
There are a lot of different functional relations between $Q$ - sums. Of special interest are those which can help us reduce powers of spectral parameter multiplying them. First of such relations is obtained by considering $B_S\left[\langle Q|w,W\rangle \right]$ in two different ways: using elementary operations for shifts of $Q$ - sums or the rule \eqref{eq:BSQ} for individual Baxter polynomials in the $Q$ - sum. Next, we can use identities 
\begin{align}
\frac{u-i/2}{u-i/2}\langle Q|w,W\rangle =& \langle Q|w,W\rangle\, , \\
\frac{u+i/2}{u+i/2}\langle Q|w,W\rangle =& \langle Q|w,W\rangle\,
\end{align} 
and use elementary operations for partial fractions to reduce their left hand sides. Finally, there are additional functional relations given by Eq.\eqref{umultVQ}. We will refer  to all these functional relations as $uQ$ - rules.

\section{Solution of Baxter equations at fixed spins}\label{fixedspins}

The aim of this Appendix is to present improved\footnote{They are more closed formula compared to \cite{VolinPerturbativeSolution}.} formula for the solution of Baxter equations at fixed integer values. As was already mentioned in the main body of the paper both inhomogeneous Baxter equations we need to solve in the case of $\N=4$ SYM reduce to the solution of the following Baxter equation:
\begin{equation}
(u+\frac{i}{2})^{2}q(u+i)+(u-\frac{i}{2})^{2}q(u-i)-(2u^{2}-\frac{1}{2}-S(S+1))q(u)=V(u)\label{eq:Baxter-equation-appendix}
\end{equation}
The first homogeneous solution  is given by (\ref{eq:Baxter-polynomial-def}).
To find the second homogeneous solution and solution of nonhomogeneous
equation we follow our procedure in the case of ABJM model \cite{uspaceTMF,ABJMtwist1}. First, using the anzats $q(u)=Q_{S}b^{[1]}$ we rewrite Baxter equation \eqref{eq:Baxter-equation-appendix} as
\begin{equation}
\nabla\left(u^{2}Q_{S}^{[-1]}Q_{S}^{[1]}\nabla b\right)=Q_{S}^{[1]}V^{[1]}
\end{equation}
so that
\begin{equation}
\nabla b=\frac{1}{u^{2}Q_{S}^{[1]}Q_{S}^{[-1]}}\Psi\left(Q_{S}^{[1]}V^{[1]}\right)
\end{equation}
and to find second solution we need to solve the following equation:
\begin{equation}
b(u)-b(u+i)=\frac{1}{u^{2}Q_{S}^{[1]}Q_{S}^{[-1]}}
\end{equation}
Using empirically found relation for even spin values  $S$
\begin{equation}
\frac{1}{u^{2}Q_{S}^{[1]}Q_{S}^{[-1]}}=\frac{1}{u^{2}}+2\sum_{n=0}^{S-1}((-1)^{n}-1)\left(\frac{1}{n-S}+\frac{1}{n+S+1}\right)\left(H_{1}(n)-H_{1}(S)\right)\left\{ \frac{Q_{n}^{[-1]}}{Q_{S}^{[-1]}}-\frac{Q_{n}^{[1]}}{Q_{S}^{[1]}}\right\} .\label{eq:secondsol-relation}
\end{equation}
it is easy to find the expression for $b(u)$:
\begin{equation}
b(u)=\eta_{2}(u)+2\sum_{n=0}^{S-1}((-1)^{n}-1)\left(\frac{1}{n-S}+\frac{1}{n+S+1}\right)\left(H_{1}(n)-H_{1}(S)\right)\frac{Q_{n}^{[-1]}}{Q_{S}^{[-1]}}.
\end{equation}
Finally, the second homogeneous solution is given by\footnote{See Appendix \ref{Hurwitz-functions} for definition of $\xi_2(u)$.}
\begin{equation}
Z_{S}(u)=Q_{S}(u)\xi_{2}(u)+P_{S}(u)\:,
\end{equation}
where
\begin{align}
P_{S}(u) & =2\sum_{k=0}^{S-1}((-1)^{k}-1)\left(\frac{1}{k-S}+\frac{1}{S+k+1}\right)\left(H_{1}(k)-H_{1}(S)\right)Q_{k}(u)\nonumber \\
& =2\sum_{k=1}^{S}(1-(-1)^{k})\left(\frac{1}{2S-k+1}-\frac{1}{k}\right)Q_{S-k}(u)\sum_{n=0}^{k-1}\frac{1}{S-n}\,.
\end{align}
To find nonhomogeneous solution we start from formal solution 
\begin{equation}
b(u)=\Psi\left(\frac{1}{u^{2}Q_{S}^{[1]}Q_{S}^{[-1]}}\Psi\left(Q_{S}^{[1]}V^{[1]}\right)\right)
\end{equation}
and use (\ref{eq:secondsol-relation}) together with easy to derive
relation $\nabla f^{[-1]}\Psi g^{[1]}=\nabla\left(f\Psi g\right)^{[-1]}-(fg)^{[-1]}$
to find
\begin{equation}
q_{nonhom}(u)=Q_{S}\Psi\left(\frac{1}{(u+i/2)^{2}}\Psi\left(Q_{S}^{[2]}V^{[2]}\right)\right)+P_{S}\Psi\left(Q_{S}V\right)-Q_{S}\Psi\left(P_{S}V\right)\,.
\end{equation}

\section{$F_S$ images}\label{F-images}

This Appendix contains expressions for the minimal set of $F_S$ - images. The other required images are then their simple sequences. First, we have
\begin{align}
F_S\left[Q_{S+k}\right] \overset{k\neq 0}{=}& -\frac{1}{k(2S+k+1)}Q_{S+k}\, , \\
F_S\left[
\langle Q|w_1,W\rangle
\right] =& \langle Q|\frac{w_1(\j)}{\j (2S-\j+1)}, W\rangle\, , \\
F_S\left[Q_S\right] =& \frac{i}{2S+1}Q_S\xi_1 -\frac{1}{2s+1}\left\{
\langle Q|\frac{1+(-1)^\j}{\j}\rangle - \langle Q|\frac{1+(-1)^\j}{2S-\j+1}\rangle
\right\}\, ,
\end{align}
\begin{multline}
F_S\left[
Q_{S+k}\xi_{a,A}
\right] \overset{k\neq 0}{=} -\frac{1}{k(2S+k+1)}Q_{S+k}\xi_a,A \\
-\frac{1}{k(2S+k+1)}F_S\left[
\frac{Q_{S+k}^{[2]}}{(u+i/2)^{a-2}}\xi_A^{[2]} - \frac{Q_{S+k}^{[-2]}}{(u-i/2)^{a-2}}\xi_A
\right]\, ,
\end{multline}
\begin{multline}
F_S\left[
\xi_{a,A}\langle Q|w,W\rangle
\right] = \xi_{a,A}\langle Q|\frac{w(\j)}{\j (2S-\j+1)},W\rangle\, + \\
F_S\left[
\frac{\xi_A^{[2]}}{(u+i/2)^{a-2}}\langle Q|\frac{w(\j)}{\j (2S-\j+1)}, W\rangle^{[2]}
-\frac{\xi_A}{(u-i/2)^{a-2}}\langle Q|\frac{w(\j)}{\j (2S-\j+1)}, W\rangle^{[-2]}
\right]\, ,
\end{multline}
\begin{multline}
F_S\left[
Q_S\xi_{a,A} 
\right] = \frac{i}{2S+1}Q_S\xi_{1,a,A} - \frac{i}{2S+1}F_S\left[
\frac{Q_S^{[2]}}{(u+i/2)^{a-1}}\xi_A^{[2]} 
\right] \\
- F_S\left[
\xi_{a,A}\langle Q|1+(-1)^\j\rangle
\right] + \frac{2}{2S+1}F_S\left[
\xi_{a,A}\langle Q| \j(1+(-1)^\j)\rangle
\right]\, .
\end{multline}
Next, introducing shortcut notation 
\begin{multline}
\Ft_S \left[
\xi_A\cdot h
\right] = \xi_A F_S [h] - F_S\left[
(u+i/2)^2(\xi_A^{[2]}-\xi_A) F_S[h]^{[2]}
\right] \\
- F_S\left[
(u-i/2)^2(\xi_A^{[-2]}-\xi_A) F_S[h]^{[-2]}
\right]
\end{multline}
we have
\begin{equation}
F_S\left[
u Q_S\xi_A
\right] = \Ft_S \left[\xi_A u Q_S\right]\, ,
\end{equation}
\begin{multline}
F_S\left[
u^2 Q_S\xi_A
\right] = \frac{2S^4+4S^3+2S^2-1}{4(2S+3)(2S-1)}F_S\left[Q_S\xi_A\right] \\
+ \Ft\left[
\xi_A \left(
u^2 Q_S - \frac{2S^4+4S^3+2S^2-1}{4(2S+3)(2S-1)}Q_S
\right)
\right]\, , 
\end{multline}
\begin{align}
F_S\left[
u^3 Q_S\xi_A
\right] =& \Ft\left[\xi_A u^3 Q_S\right]\, , \\
F_S\left[
Q_{S-1}\xi_A
\right] =& \Ft_S \left[
\xi_A Q_{S-1}
\right]\, ,
\end{align}
\begin{equation}
F_S \left[
u Q_{S-1}\xi_A
\right] = -\frac{iS^2}{2(2S-1)}F_S\left[Q_S\xi_A\right] 
+ \Ft\left[
\xi_A\left(
uQ_{S-1} + \frac{iS^2}{2(2S-1)}Q_S
\right)
\right]\, ,
\end{equation}
\begin{equation}
F_S\left[
u^2 Q_{S-1}\xi_A
\right] = \Ft_S\left[
\xi_A u^2 Q_{S-1}
\right]\, ,
\end{equation}
\begin{multline}
F_S \left[
u^3 Q_{S-1}\xi_A
\right] = -\frac{iS^2(3S^4-4S^2-6)}{8(2S+3)(2S-1)(2S-3)} F_S\left[\xi_A Q_S\right] \\
+ \Ft\left[
\xi_A\left(
u^3 Q_{S-1} + \frac{iS^2(3S^4-4S^2-6)}{8(2S+3)(2S-1)(2S-3)}Q_S
\right)
\right]\, ,
\end{multline}
\begin{multline}
F_S\left[
\frac{1}{(u+i/2)^a}\xi_{b,A}
\right] = Q_S\xi_{2,a,b,A} - F_S\left[
\xi_{a,b,A}\left(Q_S^{[-2]}-Q_S^{[2]}\right)
\right] \\
+ F_S\left[
\frac{1}{(u+i/2)^{a+b}}\xi_A^{[2]}
\right] - F_S\left[
\frac{1}{(u+i/2)^a}\left(Q_S^{[2]}-1\right)\xi_{b,A}^{[2]}
\right]\, ,
\end{multline}
\begin{equation}
F_S\left[
\frac{1}{(u+i/2)^a}
\right] = Q_S\xi_{2,a} - F_S\left[
\xi_a\left(
Q_S^{[-2]} - Q_S^{[2]}
\right) 
\right] - F_S\left[
\frac{1}{(u+i/2)^a}\left(Q_S^{[2]}-1\right)
\right]\, ,
\end{equation}
\begin{multline}
F_S\left[
\frac{1}{(u-i/2)^a}\xi_A
\right] = Q_S\xi_{a+2,A} + F_S\left[
\frac{1}{(u+i/2)^a}Q_S^{[2]}\xi_A^{[2]}
\right] - F_S\left[
\frac{1}{(u-i/2)^a}\left(Q_S^{[-2]}-1\right)\xi_A
\right]\, ,
\end{multline}
\begin{equation}
F_S\left[
\frac{1}{(u-i/2)^a}
\right] = Q_S\xi_{a+2} + F_S\left[
\frac{1}{(u+i/2)^a}Q_S^{[2]}
\right] - F_S\left[
\frac{1}{(u-i/2)^a}\left(
Q_S^{[-2]}-1
\right)
\right]\, 
\end{equation}
and ($f(n)=2S+n$):
\begin{align}
F_S\left[
u\langle Q|w_1,W\rangle
\right] =& \frac{u}{(S+2)(S-1)}w_1(S)|W\rangle_S \nonumber \\
& + \frac{i}{2f(1)f(3)}\left\{
W_1^{\{-1,-1\}}\left[
\frac{1}{\j}
\right] + W_1^{\{-1,1\}}\left[
\frac{1}{2S-\j+1}
\right]
\right\} \nonumber \\
& -\frac{i}{2f(1)f(-1)}\left\{
W_1^{\{-1,-1\}}\left[\frac{1}{2S-\j+1}\right]+
W_1^{\{-1,1\}}\left[\frac{1}{\j}\right]
\right\} \nonumber \\
& + \frac{u}{f(1)}\left\{
W_1^{\{0,1\}}\left[
\frac{1}{\j}
\right] + W_1^{\{0,1\}}\left[
\frac{1}{2S-\j+1}
\right]
\right\}\, ,
\end{align}
where $W_1^{\{n,m\}}[w] = \langle Q^{\{n\}}w(\j-m)|w_1,W\rangle_{\{1,S-1\}}$\, , 
\begin{align}
F_S\left[
u\langle Q|w_1\rangle 
\right] =& w_1(1) F_S\left[u\langle Q|1\rangle\right] \nonumber \\
& +\frac{i}{2f(1)f(3)}\left\{
W_2^{\{-1,-1\}}\left[\frac{1}{\j}\right] + 
W_2^{\{-1,1\}}\left[\frac{1}{2s-\j+1}\right]\right\} \nonumber \\
& -\frac{i}{2f(1)f(-1)}\left\{
W_2^{\{-1,-1\}}\left[\frac{1}{2S-\j+1}\right] + 
W_2^{\{-1,1\}}\left[\frac{1}{\j}\right]
\right\} \nonumber \\
& + \frac{u}{f(1)}\left\{
W_2^{\{0,1\}}\left[
\frac{1}{\j}
\right] + W_2^{\{0,1\}}\left[
\frac{1}{2S-\j+1}
\right]
\right\}\, ,
\end{align}
where $W_2^{\{n,m\}}[w] = \langle Q^{\{n\}}w(\j-m)|w_1(\j)-w_1(1)\rangle$\, ,
\begin{align}
F_S\left[
u\langle Q|1\rangle
\right] =& \frac{S^2}{2f(1)f(-1)}\left\{
Q_S\xi_1 + i\langle Q|\frac{(-1)^\j}{\j}\rangle - i\langle Q|\frac{(-1)^\j}{2S-\j+1}\rangle
\right\} \nonumber \\
& + \frac{i}{2f(1)f(3)}\left\{
(1+S)^2\langle Q|\frac{1}{\j}\rangle - \frac{2S^3+3S^2+1}{f(-1)}\langle Q|\frac{1}{2S-\j+1}\rangle
\right\} \nonumber \\
& + \frac{1}{f(-1)f(3)}\left\{
-\frac{iSf(-3)}{4}Q_{S-1} + \frac{2iS^2}{f(-1)}Q_S + 4u\langle Q|1\rangle
\right\}\, ,
\end{align}
\begin{align}
F_S\left[
u\langle Q|(-1)^\j\rangle
\right] =& -\frac{S^2}{2f(1)f(-1)}\left\{
Q_S\xi_1 + i\langle Q|\frac{1}{\j}\rangle - i\langle Q|\frac{1}{2S-\j+1}\rangle
\right\} \nonumber \\ &
-\frac{i}{2f(1)f(3)}\left\{
(1+S)^2\langle Q|\frac{(-1)^\j}{\j}\rangle - \frac{2S^3+3S^2+1}{f(-1)}\langle Q|\frac{(-1)^\j}{2S-\j+1}\rangle
\right\} \nonumber \\ &
+ \frac{1}{f(-1)f(3)}\left\{
-\frac{iSf(-3)}{4}Q_{S-1} - \frac{2iS^2}{f(-1)}Q_S +4u\langle Q|(-1)^\j\rangle
\right\}\, ,
\end{align}
\begin{align}
F_S\left[
u^2\langle Q|w_1,W\rangle
\right] =& \sum_{k=1}^2F_S\left[u^2Q_{S-k}\right]w_1(k)|W\rangle_k + \sum_{k=S-1}^SF_S\left[u^2Q_{S-k}\right]w_1(k)|W\rangle_k \nonumber \\ &
+ \frac{u^2}{f(1)}\left\{
W_3^{\{0,2\}}\left[\frac{1}{\j}\right] +
W_3^{\{0,2\}}\left[\frac{1}{2S-\j+1}\right]
\right\} \nonumber \\ &
+ \frac{iu}{2f(1)f(5)}\left\{
W_3^{\{-1,-2\}}\left[\frac{1}{\j}\right] + 
W_3^{\{-1,2\}}\left[\frac{1}{2S-\j+1}\right]
\right\} \nonumber \\ & - \frac{iu}{2f(1)f(-3)}\left\{
W_3^{\{-1,-2\}}\left[\frac{1}{2S-\j+1}\right] + 
W_3^{\{-1,2\}}\left[\frac{1}{\j}\right]
\right\} \nonumber \\ &
+ \frac{2S^4+4S^3+2S^2-1}{4f(-1)f(1)f(3)}\left\{
W_3^{\{0,0\}}\left[\frac{1}{\j}\right] + 
W_3^{\{0,0\}}\left[\frac{1}{2S-\j+1}\right]
\right\} \nonumber \\ &
-\frac{(S+2)^4}{4f(1)f(3)f(5)}\left\{
W_3^{\{0,-2\}}\left[
\frac{1}{\j}
\right] + W_3^{\{0,2\}}\left[
\frac{1}{2S-\j+1}
\right] 
\right\} \nonumber \\ &
-\frac{(S-1)^4}{4f(1)f(-3)f(-1)}\left\{
W_3^{\{0,-2\}}\left[
\frac{1}{2S-\j+1}
\right] + W_3^{\{0,2\}}\left[
\frac{1}{\j}
\right]
\right\}\, ,
\end{align}
where $W_3^{\{n,m\}}[w]=\langle Q^{\{n\}}w(\j-m)|w_1,W\rangle_{3,S-2}$\, ,
\begin{align}
F_S\left[
u^3\langle Q|w_1, W\rangle
\right] =& \sum_{k=1}^3 F_S\left[u^3 Q_{S-k}\right] w_1(k) |W\rangle_k +
\sum_{k=S-2}^S F_S\left[u^3 Q_{S-k}\right] w_1(k) |W\rangle_k \nonumber \\ &
+ \frac{u^3}{f(1)}\left\{
W_4^{\{0,3\}}\left[
\frac{1}{\j}
\right] + W_4^{\{0,3\}}\left[
\frac{1}{2S-\j+1}
\right] 
\right\} \nonumber \\ &
-\frac{iu^2}{2f(1)f(-5)}\left\{
W_4^{\{-1,3\}}\left[
\frac{1}{\j}
\right] + W_4^{\{-1,-3\}}\left[
\frac{1}{2S-\j+1}
\right] 
\right\} \nonumber \\ &
+ \frac{iu^2}{f(1)f(7)}\left\{
W_4^{\{-1,3\}}\left[\frac{1}{2S-\j+1}\right]
+ W_4^{\{-1,-3\}}\left[\frac{1}{\j}\right]
\right\} \nonumber \\ &
-\frac{u}{4f(1)}\left\{
\frac{(S+3)^4}{f(5)f(7)}W_4^{\{0,-3\}}\left[
\frac{1}{\j}
\right] + \frac{(S-2)^4}{f(-5)f(-3)}W_4^{\{0,-3\}}\left[
\frac{1}{2S-\j+1}
\right]
\right\} \nonumber \\ &
+ \frac{u (3S^4-4S^2-6)}{4f(-3)f(1)f(3)}W_4^{\{0,1\}}\left[
\frac{1}{\j}
\right] \nonumber \\ &
+ \frac{u (3S^4+12S^3+14S^2+4S-7)}{4f(-1)f(1)f(5)}W_4^{\{0,1\}}\left[
\frac{1}{2S-\j+1}
\right] \nonumber \\ &
-\frac{u(2S^4-12S^3+26S^2-24S+7)}{4f(-5)f(-1)f(1)}W_4^{\{0,3\}}\left[
\frac{1}{\j}
\right] \nonumber \\ &
-\frac{u(2S^4+20S^3+74S^2+120S+71)}{4f(1)f(3)f(7)}W_4^{\{0,3\}}\left[
\frac{1}{2S-\j+1}
\right] \nonumber \\ &
+\frac{i(S-1)^4}{8f(-5)f(-3)f(-1)f(1)}\left\{
W_4^{\{-1,-3\}}\left[
\frac{1}{2S-\j+1}
\right] + W_4^{\{-1,3\}}\left[
\frac{1}{\j}
\right]
\right\} \nonumber \\ &
+ \frac{i(3S^4+12S^3+14S^2+4S-7)}{8f(-1)f(1)f(3)f(5)}\left\{
W_4^{\{-1,-1\}}\left[
\frac{1}{\j}
\right] + W_4^{\{-1,1\}}\left[
\frac{1}{2S-\j+1}
\right] 
\right\} \nonumber \\ &
-\frac{i(3S^4-4S^2-6)}{8f(-3)f(-1)f(1)f(3)}\left\{
W_4^{\{-1,-1\}}\left[
\frac{1}{2S-\j+1}
\right] + W_4^{\{-1,1\}}\left[
\frac{1}{\j}
\right]
\right\} \nonumber \\ &
-\frac{i(S+2)^4}{8f(1)f(3)f(5)f(7)}\left\{
W_4^{\{-1,-3\}}\left[
\frac{1}{\j}
\right] + W_4^{\{-1,3\}}\left[
\frac{1}{2S-\j+1}
\right]
\right\}\, , 
\end{align}
where $W_4^{\{n,m\}}[w] = \langle Q^{\{n\}}w(\j-m)|w_1,W\rangle_{\{4,S-3\}}$. In addition, if $p(u)$ is a polynomial we have 
\begin{equation}
F_S\left[\xi_A p(u)\right] = \Ft_S\left[\xi_A p(u)\right]\, .
\end{equation}
and ($g$ is any expression)
\begin{equation}
F\left[\PP_k (u) g\right] = \PP_k (u) F[g]\, .
\end{equation}

\section{$\Psi$ images}\label{Psi-images}

Here we have gathers expressions for a minimal set of $\Psi$ - images. The other required images are their simple sequences. We have 
\begin{equation}
\Psi\left[
Q_S
\right] = \frac{1}{2(2S+1)}\left\{
(S+1)Q_{S+1} + (2S+1)Q_S + S Q_{S-1}
\right\}\, ,
\end{equation}
\begin{align}
\Psi\left[
\langle Q|w_1, W\rangle
\right] =& \frac{1}{2}\langle Q|w_1,W\rangle + i u \langle Q|\frac{w_1(\j)}{S-\j+1}, W\rangle\nonumber \\ &
+ \frac{1}{2}\left\{
W_5^{\{-1,-1\}}\left[
\frac{1}{S-\j+1}
\right] - W_5^{\{-1,0\}}\left[
\frac{1}{S-\j+1}
\right]
\right\}\, ,
\end{align}
where $W_5^{\{n,m\}}[w] = \langle Q^{\{n\}}w(\j-m)|w_1, W\rangle$.
\begin{align}
\Psi\left[
u\langle Q|w_1, W\rangle
\right] = & \sum_{k=S-1}^S\Psi\left[uQ_{S-k}\right]w_1(k)|W\rangle_k
+ \frac{i(12u^2-1)}{12}W_6^{\{0,1\}}\left[
\frac{1}{S-\j+1}
\right] \nonumber \\ &
+ \frac{u}{6}\left\{
W_6^{\{-1,-2\}}\left[
\frac{1}{S-\j+1}  
\right] - W_6^{\{-1,1\}}\left[
\frac{1}{S-\j+1}  
\right]
\right\} \nonumber \\ &
+\frac{u}{2}W_6^{\{0,0\}}[1] + \frac{i}{12}W_6^{\{0,-2\}}\left[
\frac{1}{S-\j+1}
\right]\, ,
\end{align}
where $W_6^{\{n,m\}}[w] = \langle Q^{\{n\}}w(\j-m)|w_1, W\rangle_{\{1,S-2\}}$. Next, images with $\xi_A$ - functions have the form \cite{VolinPerturbativeSolution} ($\nabla f = f - f^{[2]}$):
\begin{equation}
\Psi\left[
\xi_A\cdot h
\right] = \xi_A\Psi [h] - \Psi\left[
\left(\nabla\xi_A\right)^{[-2]}\Psi [h]
\right]^{[2]}\, ,
\end{equation}
where $h$ is some polynomial build from Baxter polynomials and their sums. In addition we also used the following images \cite{VolinPerturbativeSolution}:
\begin{align}
\Psi\left[
\frac{1}{(u+i/2)^a}\xi_{b,A}\right]  =& \xi_{a,b,A} + \xi_{a+b,A}
\, , \\ 
\Psi\left[
\frac{1}{(u-i/2)^a}\xi_{b,A}
\right] =& \xi_{a,b,A} + \frac{1}{(u-i/2)^a}\xi_{b,A}\, ,\\
\Psi\left[
\frac{1}{(u+i n/2)^a}\right] =& \xi_a^{[n-1]}\, , \\
\Psi\left[
\PP_k (u) g
\right] =& \PP_k (u)\Psi [g]\, .
\end{align}

\bibliographystyle{hieeetr}
\bibliography{litr}

\end{document}